\documentclass{iopart}
\usepackage{iopams}

\usepackage[margin=3cm]{geometry} 
\usepackage[utf8]{inputenc}
\usepackage[english]{babel}

\usepackage{subcaption}
\usepackage{caption} 
\usepackage{graphicx}
\usepackage[export]{adjustbox}
\usepackage{xcolor}
\usepackage{pgfplots} 
\usepackage{tikz}  
\usepackage{tabu} 

\usepackage{dsfont} 
\usepackage{esint}     
\usepackage{bm}
\usepackage{bbold}
\usepackage{braket} 
\usepackage{hhline} 
\usepackage{blkarray} 
\usepackage{multirow} 
\usepackage{esvect}

\usepackage{hyperref}
\usepackage{ifthen} 
\usepackage{verbatim}
\usepackage[toc,page]{appendix}

\newcommand{\sech}{\mathop{\mathrm{sech}}}
\newcommand{\diag}{\mathop{\mathrm{diag}}}
\DeclareMathDelimiter{\lVert}
  {\mathopen}{symbols}{"6B}{largesymbols}{"0D}
\DeclareMathDelimiter{\rVert}
  {\mathclose}{symbols}{"6B}{largesymbols}{"0D}
\newcommand\norm[1]{\lVert#1\rVert}
\renewcommand\bra[1]{{\langle{#1}|}}
\makeatletter
\renewcommand\ket[1]{%
  \@ifnextchar\bra{\k@t{#1}\!}{\k@t{#1}}%
}
\newcommand\k@t[1]{{|{#1}\rangle}}

\newcommand{\ketbra}[2]{\ket{#1}\bra{#2}}

\makeatletter
\DeclareRobustCommand{\rvdots}{%
  \vbox{
    \baselineskip4\p@\lineskiplimit\z@
    \kern-\p@
    \hbox{.}\hbox{.}\hbox{.}
  }}
\makeatother

\usetikzlibrary{positioning}
\usetikzlibrary{decorations.pathreplacing,calc}

\newcommand{\detector}[3]{
    \begin{scope}[]
    \fill[#1] (#2,#3) -- (#2,#3) arc(-90:90:0.2) --cycle;
    \end{scope}
    }

    \usetikzlibrary{chains}
    
\renewcommand{\vec}[1]{\boldsymbol{#1}}
\newcommand{\w}{\omega}
\newcommand{\sfrac}[2]{{\textstyle{\frac{ #1}{#2}}}}

\pgfplotsset{compat=1.16}

\begin{document}

\title[A general framework for multimode Gaussian quantum optics and photo-detection]
{A general framework for multimode Gaussian quantum optics and photo-detection: application to Hong--Ou--Mandel interference with filtered heralded single photon sources}
\author{Oliver F. Thomas$^{1,2}$, Will McCutcheon$^{1,3}$ and Dara P. S. McCutcheon$^1$}
\address{$^1$Quantum Engineering Technology Labs, H. H. Wills Physics Laboratory and Department of Electrical and Electronic Engineering, University of Bristol, BS8 1FD, United Kingdom \\ $^2$Quantum Engineering Centre for Doctoral Training, H. H. Wills Physics Laboratory and Department of Electrical and Electronic Engineering, University of Bristol, Bristol, UK \\ $^3$BBQLabs, Institute of Photonics and Quantum Sciences, Heriot-Watt University, Edinburgh, EH14 4AS, United Kingdom}

\date{\today}

\begin{abstract}
    The challenging requirements of large scale quantum information processing using parametric heralded single photon sources involves maximising the interference visibility whilst maintaining an acceptable photon generation rate. By developing a general theoretical framework that allows us to include large numbers of spatial and spectral modes together with linear and non-linear optical elements, we investigate the combined effects of spectral and photon number impurity on the measured Hong--Ou--Mandel interference visibility of parametric photon sources, considering both threshold and number resolving detectors, together with the effects of spectral filtering. We find that for any degree of spectral impurity, increasing the photon generation rate necessarily decreases the interference visibility, even when using number resolving detection. While tight spectral filtering can be used to enforce spectral purity and increased interference visibility at low powers, we find that the induced photon number impurity results in a decreasing interference visibility and heralding efficiency with pump power, while the maximum generation rate is also reduced.
\end{abstract}
\section{Introduction}

Almost all tasks in optical implementations of quantum communication~\cite{duan2001long, gisin2007quantum, ursin2007entanglement}, quantum cryptography~\cite{beveratos2002single, townsend1997quantum,hughes1995quantum}, quantum sensing and quantum information processing \cite{knill2001scheme, nielsen2004optical, varnava2008good} require sources of single photons~\cite{bennett1984proceedings,fuchs1997optimal}. Although the requirements of such a source depend on the particular application~\cite{aharonovich2016solid, brendel1999pulsed}, an ideal source would produce single photons with high spectral purity, and either deterministically or with a very high efficiency. To this end, parametric non-linear processes can produce pairs of correlated photons in distinct modes, with the detection of a photon in one mode heralding the presence of a single photon in another. Such sources benefit from potentially very high photon spectral purity~\cite{paesani2020near}, and can in many cases be more readily incorporated into integrated devices with a high density and reproducibility~\cite{Wang2018,Wang2020}. Such sources, however, are not deterministic, and in order to achieve high efficiencies, it is necessary to multiplex many sources together~\cite{meany2014hybrid,ma2011experimental}. While many of the impressive purity metrics of these sources have been measured in the weak-excitation limit, an analysis of the level of multiplexing necessary to achieve a given efficiency obviously requires operation beyond this limit, and in conjunction with any effects of loss, including filtering~\cite{Bonneau2015,Francis-Jones2014}. This problem is further complicated when one also considers the types of detectors employed, as beyond the weak excitation limit photon number resolving detectors project states in a way not possible with threshold (or `bucket') detectors, which instead only indicate the presence of one or more photons.

Another route towards making a deterministic source is to instead use single quantum emitters such as semiconductor quantum dots~\cite{lodahl2015interfacing}, defect centres in crystals and 2D materials~\cite{aharonovich2016solid}, and single organic molecules~\cite{Clear2020,Rezai2018}. Using these systems, an excited state can be populated on demand, and radiative relaxation can then occur with very high internal quantum efficiency, with the emitted photon in many cases having a high spectral purity~\cite{Somaschi2016,Ding2016,uppu2020scalable}. However, there are often efficiency--purity trade-offs using such sources~\cite{Iles-Smith2017}, as it is typically necessary to strongly modify the surrounding photonic environment to boost extraction efficiencies or increase spectral purity (using e.g. cavities), which can simultaneously have detrimental effects on other figures of merit. These considerations, and the challenges involved in incorporating many such sources into a scalable platform~\cite{lodahl2015interfacing} leaves significant questions regarding their utility in information processing applications.

For these reasons, the advantages offered by parametric sources motivates a detailed study of their operation beyond the weak excitation limit. Whilst it is possible to write down the quantum state of the field produced by a parametric source in the Fock basis, and from this evaluate the heralding rate and fidelity to a pure single photon state including spectral impurity and multi-photon events~\cite{Christ2012}, it is not a straightforward matter to propagate many such multi-photon multi-modal states through subsequent optical elements and calculate measured detection probabilities. This challenge arises due to the cumbersome nature of dealing directly with large Fock states and points towards the difficulties in the concatenation of multiple systems and tracking of errors. This represents a problem even when seeking to accurately model Hong--Ou--Mandel (HOM) interference of two heralded photons on a subsequent beam-splitter, and particularly so in the presence of loss or if spectral filters are used, since strongly frequency dependent photon number states evolve non-trivially through latter devices in the system. Regarding detection, recent work on Gaussian boson sampling provides expressions for the detection probabilities using threshold detectors~\cite{quesada2018gaussian} and number resolving detectors~\cite{kruse2019detailed}, while multiphoton contributions for threshold detectors have been investigated for single-mode sources~\cite{takeoka2015full,tiedau2019scalability}, although in all cases ignoring the photon spectral properties. Spectrally multi-moded sources have been investigated but only for measurements of intensity correlation functions~\cite{christ2011probing,faruque2019estimating}, which are not in general the appropriate measurements to investigate single photon interference.

In this work we present a general framework of multimode Gaussian optics which includes arbitrary spatial and spectral degrees of freedom, arbitrary photon numbers, and readily accounts for linear optical elements such as beam-splitters and filters, as well as non-linear squeezing operations which represent parametric sources. The Gaussian formalism offers a means to efficiently model an entire system, requiring propagation of a covariance matrix of size only linear in the number of modes, followed by a threshold or photon counting detection model which extends previous results to the multi-spectral moded case. We put this new framework to use by performing a thorough evaluation of single photons heralded from a parametric photon pair source valid for all excitation powers and corresponding photon generation rates, and in which we simultaneously account for spectral impurity and the effects of loss and filtering. We assess the degree of quantum interference from a simulation of the experimentally measured HOM interference statistics between heralded photons from two sources, elucidating the quantitative and qualitative differences found using both detector types. As has been previously established, for threshold detectors we find that the HOM interference visibility decreases with increasing photon generation rate, as at higher powers multiphoton components contaminate the heralded `single photon' state~\cite{Christ2012}. When spectral impurity is included, this problem is also present when using number resolving detectors and post-selecting only one-photon events; even in the single photon subspace increasing the photon generation rate necessarily and detrimentally affects the distribution of spectral modes in the heralded (truly) single photon state. With the inclusion of spectral filtering, while the interference visibility can be made arbitrarily high in the weak excitation limit, any increases in power to improve heralding rates deteriorate both the interference visibility and heralding efficiency. These results indicate that considerable care should be taken when designing a source with targeted figures or merit, but at the same time provides a general framework to efficiently describe the larger systems for which the sources are intended, allowing for the consequences of multiple and varied source imperfections to be captured.

\section{Gaussian symplectic formalism and photon detection}

\begin{figure}[t!]
    \centering
    \includegraphics[width=0.75\textwidth]{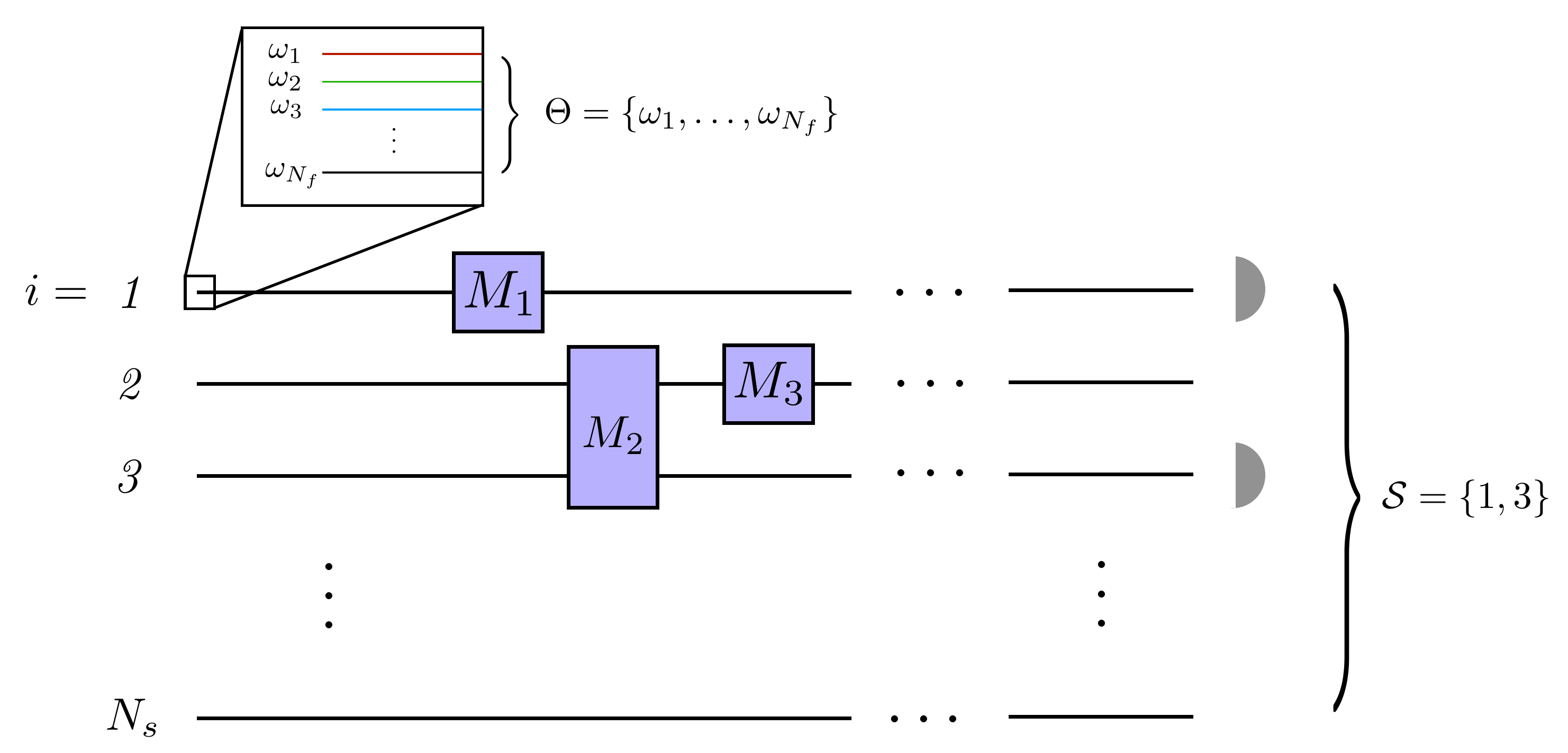}
    \caption{The general optical circuit that our presented formalism applies to consists of a collection of modes which each carry one of $N_s$ spatial labels $i$ indicating, for example, in which waveguide they are confined, and one of $N_f$ spectral labels $\w$ indicating their colour or frequency. A succession of multimode Gaussian transformations labelled $M_j$ (which may be linear or non-linear) couple the modes. Threshold or number resolving photon detection then takes place in a subset $\mathcal{S}$ (for illustrative purposes here {\it{1}} and {\it{3}}) of the spatial modes.}
    \label{fig:generalcircuit}
\end{figure}

The general scenario to which our formalism applies is depicted in Fig.~{\ref{fig:generalcircuit}}, and consists of a collection of optical modes, which can be excited, coupled, and the states of which can be detected. As we are interested in accurately describing the spectral properties of photons, we take particular care to distinguish between spatial and spectral mode labels. A spatial mode label indicates where (e.g. in which waveguide) a mode is defined, while a spectral label refers to the frequency or wavelength of that mode. To accommodate both of these properties, we label the creation and annihilation operator of a mode with two indices, generically $i$ and $\w$, referring to spatial and spectral degrees of freedom, respectively. We list all of our mode operators in a vector, which takes the form
\begin{equation}
\hat{\vec{A}}=({\hat{\vec a}}_{1},\dots,{\hat{\vec a}}_{N_s},{\hat{\vec a}}^{\dagger}_{1},\dots,{\hat{\vec a}}^{\dagger}_{N_s})^\top,
\label{Aspectralandspatial}
\end{equation}
where $N_s$ is the number of spatial mode labels, and for each spatial mode label $i$ we have the vector $\hat{\vec{a}}_i = (\hat{a}_{i\w_1},\dots,\hat{a}_{i \w_{N_f}})^\top$, with $\hat{a}_{i\w}$ being the annihilation operator for a mode with spatial label $i$ and spectral label $\w$. The number of spectral (equivalently frequency) modes is $N_f$, giving a total number of modes $N=N_s N_f$. The mode operators satisfy the usual commutation relations $[\hat{a}_{i\w},\hat{a}^{\dagger}_{j\omega'}]=\delta_{ij}\delta_{\w\w'}$.

A variety of candidates exist for parametric photon sources, including free-space systems~\cite{Valencia2020}, cavity-based systems~\cite{McCutcheon2020}, in-line or wave-guided systems~\cite{Quesada2019}, and inter-modal phase-matched systems~\cite{paesani2020near}. Crucially, each shares the common feature that their dynamics are generated by Hamiltonians which are at most quadratic in the quantum field mode operators. While the underlying electromagnetic nonlinearity may be cubic or quartic, in all cases mentioned above only two of the field operators in the Hamiltonian are treated quantum mechanically, and the remaining bright fields contribute as time-dependent scalars. Consequently, once the dynamics are solved, the solutions are Gaussian transformations, which are described by an effective Hamiltonian $\hat{H}$ which is quadratic in the field mode operators of interest, and which can in all generality be written 
\begin{equation}
\hat{H} = \hat{\vec{A}}^\dagger \mathbb{H} \hat{\vec{A}},
\label{BilinearH}
\end{equation}
where $\mathbb{H}$ is a matrix of scalar coefficients which may be a discrete approximation to a continuous function under appropriate regularity conditions~\cite{Quesada2019}. The Hamiltonian has corresponding unitary time evolution operator $\hat{U}=\exp \small[- i \hat{H} ]$, and using $\hat{H}=\hat{H}^{\dagger}$ and the basic commutation relations for the creation and annihilation operators one can show~\cite{adesso2014continuous}
\begin{equation}
\hat{U}^{\dagger} \hat{\vec{A}} \hat{U}=M\hat{\vec{A}};\qquad
M=\exp[-2i K \mathbb{H}],
\label{MfromH}
\end{equation}
and where 
\begin{equation}
K=\left(
    \begin{array}{cc}
        1 & 0 \\
        0 & -1
    \end{array}\right)\otimes\mathds{1}_N,
\end{equation}
with $\mathds{1}_N$ the $N$-dimensional identity matrix. Since the commutation relation between the mode operators is preserved under a unitary transformation we have 
\begin{equation}
MKM^{\dagger}=K,
\label{SymplecticCondition}
\end{equation}
or equivalently $M^{\dagger}K=KM^{-1}$, and which defines $M$ as a linear symplectic matrix~\footnote{The symplectic condition is usually written 
$M \Omega M^\top = \Omega$, 
with $\Omega=\left(\begin{array}{cc} 0 & 1 \\ -1 & 0 \end{array}\right)\otimes \mathds{1}_N$,
which is equivalent to Eq.~({\ref{SymplecticCondition}}) if 
$M\to L M L^{\dagger}$ with 
$L=\frac{1}{\sqrt{2}}\left(\begin{array}{cc} 1 & i \\ 1 & -i \end{array}\right)\otimes \mathds{1}_N$.}. 
Eq.~({\ref{MfromH}}) provides us with a way in which to use the underlying Hamiltonian parameters to propagate the collection of mode operators in the Heisenberg picture.

\subsection{Gaussian states and transformations}

Rather than working directly with the quantum state of the optical modes $\hat{\rho}$, we instead consider the corresponding characteristic functions~\cite{adesso2014continuous}. We define the $s$-ordered characteristic function as 
\begin{equation}
\chi_{\hat{\rho}}^{(s)}(\vec{\Lambda})=\mathrm{Tr}[\hat{\rho} \hat D(\vec{\Lambda})]\exp[\sfrac{1}{4} s |\vec{\Lambda}|^2],
\end{equation}
where $\hat D(\vec{\Lambda})=\smash{\exp \big[\hat{\vec{A}}^{\dagger} K \vec{\Lambda}\big]}$ and $\vec{\Lambda}$ is a $2N$-dimensional vector, the elements of which are complex variables which describe the quantum state in an associated phase-space. As we detail in the appendix, a Gaussian state is one having a Gaussian characteristic function of the complex variables $\vec{\Lambda}$, and is therefore uniquely determined by a $2N\times 2N$ matrix of scalar coefficients $\sigma$ which is known as the covariance matrix, and a $2N$-dimensional vector $\vec{d}$ called the displacement vector. In this work we will exclusively consider states for which $\vec{d}=0$. If a Gaussian state undergoes a linear symplectic transformation as in Eq.~({\ref{MfromH}}), it follows that its covariance matrix transforms as 
\begin{equation}
\sigma\to\sigma'=M\sigma M^{\dagger}.
\label{sigmaTransformation}
\end{equation}
By noting that the vacuum has corresponding covariance matrix $\mathds{1}_{2N}$, we see that Eq.~({\ref{sigmaTransformation}}) allows us to construct a description of a quantum state generated by successive symplectic transformations.

In what follows we show that it is possible to model both threshold (`bucket') and number resolving detectors using only projections of a state onto the vacuum for different subsets of the modes. For this reason, let us therefore consider a subset of the total $N$ modes, which we label $S$. We are then interested in the probability associated with the projector onto the vacuum for all modes in $S$, which we label $\ketbra{\mathrm{vac}}{\mathrm{vac}}_S$. As shown in the appendix, the projection of a general Gaussian state described by a covariance matrix $\sigma$ onto this state can be found to be~\cite{takeoka2015full}
\begin{equation}
P_{\mathrm{off}}(S)=\mathrm{Tr}[\rho \ketbra{\mathrm{vac}}{\mathrm{vac}}_S] = \left(\mathrm{det}\left[(\mathds{1}_{2|S|} + \sigma_S)/2\right] \right)^{-1/2},
\label{vacExpectationValue}
\end{equation}
where $|S|$ is the number of modes in $S$ and $\sigma_S$ is the $2|S|\times 2|S|$ covariance matrix pertaining only to modes in $S$. Setting $\sigma_S=\mathds{1}_{2|S|}$ we find $P_{\mathrm{off}}(S)=1$ as expected.

\subsection{Photon detection using threshold detectors}

We now consider the probabilities associated with photon detection using threshold detectors, extending the existing results which include only multiple spectral modes~\cite{takeoka2015full} or multiple spatial modes~\cite{quesada2018gaussian}. Threshold detectors are currently the most widely used experimentally, returning a signal (or `click') with a probability which depends on the presence of one or more photons in a given spatial mode. These detectors do not typically resolve spectral degrees of freedom, and as such the relevant projection operator corresponding to a detector click in {\emph{any}} spectral mode with spatial label $i$ is 
\begin{equation}
\ketbra{\mathrm{on}}{\mathrm{on}}_{i} = \prod_\w \hat{\mathds{1}}_{i\w} - \prod_\w \ketbra{\mathrm{vac}}{\mathrm{vac}}_{i\w},
\label{onProjectorSpec}
\end{equation}
which is the projection onto all states with spatial mode label $i$ except the vacuum on all $N_f$ spectral modes. In the following we use the calligraphic notation $\mathcal{S}$ to represent the set of spatial mode labels in which the detection events take place, while $\Theta=\{\w_1,\dots,\w_{N_f}\}$ is the set of all spectral mode labels. As such, a complete set of (spatial and spectral) labels is $S=\mathcal{S}\times\Theta$. For example, we may be interested in spatial modes $\mathcal{S}=\{1,3\}$ (see Fig.~{\ref{fig:generalcircuit}}), which including spectral labels gives the set $S=\{1,3\}\times\{\w_1,\dots,\w_{N_f}\}=\{1\w_1,\dots,1\w_{N_f},3\w_1,\dots,3\w_{N_f}\}$.

Using this, the probability to detect at least one photon in each of the spatial modes $i\in \mathcal{S}$ (and with any spectral mode labels) is 
\begin{equation}
\mathcal{P}_{\mathrm{on}}(\mathcal{S}) = \mathrm{Tr}\big[\hat{\rho} \prod_{i\in \mathcal{S}}\ketbra{\mathrm{on}}{\mathrm{on}}_{i}\big] = \sum_{\mathcal{B} \in 2^\mathcal{S}} (-1)^{|\mathcal{B}|} \mathcal{P}_{\mathrm{off}}(\mathcal{B}),
\label{Pon}
\end{equation}
where $2^\mathcal{S}$ is the power set (the set of all subsets) of $\mathcal{S}$ and 
\begin{equation}
\mathcal{P}_{\mathrm{off}}(\mathcal{B})=\mathrm{Tr}\big[\hat{\rho}\prod_{i\in\mathcal{B}}\prod_\w \ketbra{\mathrm{vac}}{\mathrm{vac}}_{i\w}\big]
=P_{\mathrm{off}}(B),
\end{equation}
with $B=\mathcal{B}\times\Theta$, and for which we can use Eq.~({\ref{vacExpectationValue}}) for Gaussian states. The general form for threshold detectors allows us to also calculate joint probabilities of clicks and vacuum detection events in sets of spatial modes. For detection events in spatial modes $\mathcal{X}$ and vacuum in spatial modes $\mathcal{Y}$ we have 
\begin{equation}
\hspace{-1cm}
\mathcal{P}_{\mathrm{on},\mathrm{off}}(\mathcal{X},\mathcal{Y}) = 
\mathrm{Tr}\big[\hat{\rho} \prod_{i\in \mathcal{\mathcal{X}}}\ketbra{\mathrm{on}}{\mathrm{on}}_{i} 
\prod_{j\in \mathcal{\mathcal{Y}},\w}\ketbra{\mathrm{vac}}{\mathrm{vac}}_{j \w }\big]=
\sum_{\mathcal{B} \in 2^{\mathcal{X}}} (-1)^{|\mathcal{B}|} \mathcal{P}_{\mathrm{off}}(\mathcal{B} \cup \mathcal{Y}).
\label{Ponoff}
\end{equation}
This expression extends the previously known result for  threshold detector probabilities involving the Torontonian function in Ref.~\cite{quesada2018gaussian}. We note that the number of explicit terms in the sum in Eq.~({\ref{Pon}}) depends only on the number of {\emph{spatial}} modes in which photon(s) are detected. The number of spectral degrees of freedom increases the size of the reduced covariance matrices in Eq.~({\ref{vacExpectationValue}}). 

\subsection{Photon number resolving detection}

We now present expressions for probabilities associated with number resolving detection in spatial modes. Our method takes inspiration from the way in which pesudo-number resolving detectors can be constructed experimentally, namely by `fanning' out a mode across multiple modes and using threshold detectors. Interestingly, we find that these photon number statistics can be constructed only considering projections onto the vacuum for different combinations of modes.

In our derivation, for each spatial mode we distribute its amplitude across $m$ fictitious ancillary modes using a unitary which has equal amplitudes across its range. Doing so creates an equal superposition of the input state diluted by the vacuum over $m$ modes, and at the end of each we conceptually place a threshold detector. Provided that the number of modes $m$ is much larger than the number of photons present in the initial state described by covariance matrix $\sigma_S$, we can assume that each of the $m$ diluted modes contains at most $1$ photon. In this way the probability for a total number of threshold detector clicks, which we label $k$, gives an approximation to the probability to detect $n$ photons in the initial state described by $\sigma_S$. Experimentally this leads to a correspondence between the number of ancillary modes used and the number of photons which can be detected accurately. In our case the ancillary modes are conceptual, allowing us to take the limit $m\to\infty$ and give exact expressions for the probability to detect a fixed number of photons $n$.

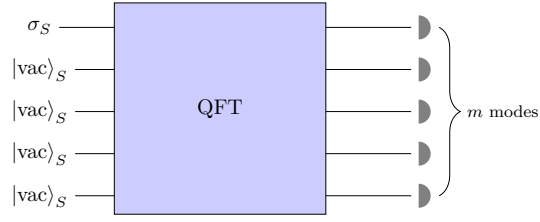
\begin{figure}[t!]
    \centering
    \scalebox{0.8}{%
    \begin{tikzpicture}
    \def\detdist{6.3cm} 
    \tikzstyle{block} = [draw, fill = blue!20, minimum size = 10em]
    \tikzstyle{smblock} = [draw, fill = blue!20, minimum size = 1 em]
    \node at (0, -0.7 * 0) (a0) {$\sigma_{S}$};
    \node[right of = a0, node distance = \detdist] (det0) {}; 
    \draw (a0) -- (det0);
    \foreach \y in {1,...,4}
    {
        \node at (0, -0.7 * \y) (a\y) {$\ket{\mathrm{vac}}_S$};
        \node[right of = a\y, node distance = \detdist] (det\y) {};
        \draw (a\y) -- (det\y);
        \detector{gray}{6.3}{-0.7 * \y -0.2};
    };
    \node[block] at (3, -1.35) (sq) {QFT};
    \detector{gray}{6.3}{-0.2};
    \draw [decorate,decoration={brace,amplitude=10pt,raise=4pt},yshift=0pt] (6.5,0) -- (6.5,-0.7*4) node [black,midway,xshift=1.2cm] {\footnotesize $m$ modes};
    \end{tikzpicture}}
    \caption{Conceptual fan-out circuit used to count the number of photons in an arbitrary Gaussian state $\sigma_S$ using threshold detectors.}
    \label{fig:qftfanout}
\end{figure}

We derive our expressions using the following steps, in part inspired by Ref.~\cite{paul1996photon}. 

\begin{enumerate}
    \item We consider a subsystem of our total Gaussian state, which is also Gaussian and described by a $2|S|\times 2|S|$ covariance matrix $\sigma_S$, which may contain both spatial and spectral degrees of freedom. As depicted in Fig.{\ref{fig:qftfanout}}, we fan this state out into $m$ supermodes (which may each contain spatial and spectral degrees of freedom), and we label the set of supermodes with boldface calligraphic symbols, i.e. $\bm{\mathcal{S}}=\{1,\dots,m\}$. The complete set of mode labels is then given by $\bm{\mathcal{S}}\times \mathcal{S}\times\Theta=\bm{\mathcal{S}}\times S$, of which there are in total $|\bm{\mathcal{S}}||S|=m|S|$. All $|S|(m-1)$ ancillary modes are initially in the vacuum state. The total initial covariance matrix is written $\sigma_{\bm{\mathcal{S}}\times S}=\sigma_S \oplus \mathds{1}_{2|S|(m-1)}$. The fanning out transformation is achieved using the quantum Fourier transform (QFT) acting on the supermodes, although the specific unitary is unimportant. The QFT evenly distributes initial amplitude in $\sigma_S$, and has the matrix elements 
    $(U_{\mathrm{QFT}})_{nn'} = \exp[2 \pi i n n'/m]/\sqrt{m}$. 
    The covariance matrix following the action of the QFT is given by 
\begin{equation}
\label{eq:fanout}
\hspace{-1.5cm}
\eqalign{
\sigma_{\bm{\mathcal{S}}\times S}'&= U_{\mathrm{QFT}} \otimes \mathds{1}_{2|S|} \left( \begin{array}{ccc} 
  \tilde \sigma_S + \mathds{1}_{2|S|} & \dots & 0\\ 
 \vdots & \ddots & \vdots \\ 
 0 & \dots & 0 + \mathds{1}_{2|S|}
\end{array} \right) U^\dagger_{\mathrm{QFT}} \otimes \mathds{1}_{2|S|}, \\
&= 
\frac{1}{m}
\left( \begin{array}{ccc}
        \tilde \sigma_{S} & \dots & \tilde \sigma_S \\
        \vdots & \ddots & \vdots \\
        \tilde \sigma_S & \dots & \tilde \sigma_S
    \end{array}\right)
    + 
 \left( \begin{array}{ccc}
      \mathds{1}_{2|S|} & \dots & 0 \\
        \vdots & \ddots & \vdots \\
        0 & \dots & \mathds{1}_{2|S|} 
    \end{array}\right)  
    = \frac{1}{m} \vv{\tilde \sigma_S} (\vv{\mathds{1}_{2|S|}})^\top + \mathds{1}_{2m|S|} }
   ,
\label{sigmaSp}
\end{equation}
where we have defined $\tilde{\sigma}_S=\sigma_S-\mathds{1}_{2|S|}$ and $\vv{\tilde \sigma_S}=(\tilde{\sigma}_S,\dots,\tilde{\sigma}_S)^\top$ as the block-constant vector with each of its $m$ elements equal to $\tilde \sigma_S$, and $\vv{\mathds{1}_{2|S|}}$ is similarly a block-constant vector with all entries the identity matrix $\mathds{1}_{2|S|}$. We see that this procedure results in a block rank 1 form for the total covariance matrix $\sigma_{\bm{\mathcal{S}}\times S}'$.

\item We next determine the probability to detect zero clicks (the vacuum) in some subset of the supermodes $\bm{\mathcal{B}}$. The quantity of interest is 
\begin{equation}
P_{\mathrm{off}}(\bm{\mathcal{B}}\times S)=
 \left(\mathrm{det}\left[(\mathds{1}_{2|S||\bm{\mathcal{B}}|} + \sigma_{\bm{\mathcal{B}}\times S})/2\right] \right)^{-1/2},
\end{equation}
where $\sigma_{\bm{\mathcal{B}}\times S}$ is the covariance matrix pertaining to the subset of modes with supermode labels in $\bm{\mathcal{B}}$, which takes on precisely the same form as in Eq.~({\ref{sigmaSp}}), but with the constant vectors $\vv{\tilde \sigma_S}$ and $\vv{\mathds{1}_{2|S|}}$ instead having length $|\bm{\mathcal{B}}|=b$. Using the matrix determinant lemma and commutative subring properties we find we can write~\footnote{We use the identity 
$\mathrm{det}[\vv{u}\vv{v}^\top+A]=\mathrm{det}[\mathds{1}+\vv{v}^\top A^{-1} \vv{u}]\mathrm{det}[A].$}
\begin{equation}
    P_{\mathrm{off}}(\bm{\mathcal{B}}\times S) = p_{\mathrm{off}}(b)=\left(\det\big[ \mathds{1}_{2|S|} + \frac{b}{2m}\tilde\sigma_{S}\big]\right)^{-1/2} \, ,
    \label{eq:vacprojfannedout}
\end{equation}
which we see depends only on the number of supermodes in subset $b$. This result also reduces the determinant of the potentially high dimensional covariance matrix in Eq.~({\ref{sigmaSp}}) to essentially the determinant of one block $\tilde{\sigma}_{S}$ which is only $2 |S|$-dimensional, meaning we can evaluate an arbitrary number of fanned out modes without changing the complexity of the determinant.

    \item We now extend Eq.~({\ref{Ponoff}}) to write down the probability to detect at least one photon in each supermode in the subset $\bm{\mathcal{X}}$ and vacuum in all other supermodes $\bm{\mathcal{Y}}$, namely 
    \begin{equation}
        \bm{\mathcal{P}}_{\mathrm{on},\mathrm{off}}(\bm{\mathcal{X}},\bm{\mathcal{Y}})
        =\sum_{\bm{\mathcal{B}}\in 2^{\bm{\mathcal{X}}}} (-1)^{|\bm{\mathcal{B}}|} P_{\mathrm{off}}((\bm{\mathcal{B}} \cup \bm{\mathcal{Y}}) \times S).
    \end{equation}
    Eq.~({\ref{eq:vacprojfannedout}}) tells us that the terms in the summation above depend only on the size of the set of supermode labels $(\bm{\mathcal{B}} \cup \bm{\mathcal{Y}})$. We then collect terms involving subsets $\bm{\mathcal{B}}$ of the same size to write 
    \begin{equation}
        \bm{\mathcal{P}}_{\mathrm{on},\mathrm{off}}(\bm{\mathcal{X}},\bm{\mathcal{Y}})
        =\sum_{l=0}^k (-1)^l {k \choose l}p_{\mathrm{off}}(m-(k-l)),
    \end{equation}
    where $k=|\bm{\mathcal{X}}|$ is the total number of detection events. As we might expect, this probability depends only on the number of detection events, and not the particular pattern. We then sum over all patterns $\bm{\mathcal{X}}$ with fixed length $k$ to give the probability for all detection patterns with $k$ clicks as 
    \begin{equation}
    \label{PXY}
        \sum_{|\bm{\mathcal{X}}|=k}\bm{\mathcal{P}}_{\mathrm{on},\mathrm{off}}(\bm{\mathcal{X}},\bm{\mathcal{Y}})={m \choose k}\sum_{l=0}^k (-1)^l {k \choose l}p_{\mathrm{off}}(m-(k-l)).
    \end{equation}
    
    \item In the limit that the number of supermodes $m$ becomes very large, the probability that any supermode contains more than one photon becomes vanishingly small, and a  threshold detector click amounts to the detection of exactly one photon. Our expression above for $k$ clicks then becomes equal to the probability to detect exactly $n$ photons. In the limit $m \to \infty$ we have $k\to n$, we can recognise the right-hand side of Eq.~({\ref{PXY}}) as a derivative, and we find the number resolving detection probability associated with the detection of $n$ photons in the state described by covariance matrix $\sigma_S$:
\begin{equation}
\hspace{-1cm}
    P_{\mathrm{PNR}}(n) = \lim_{m\to\infty}\sum_{|\bm{\mathcal{X}}|=n}\bm{\mathcal{P}}_{\mathrm{on},\mathrm{off}}(\bm{\mathcal{X}},\bm{\mathcal{Y}})=\left. \frac{(-1)^n}{n!} \partial^n_t \det\left(\mathds{1}_{2|S|} + \frac{t}{2}\tilde\sigma_S \right)^{-1/2} \right|_{t=1} \, .
\end{equation}
    \item Similar steps can be used to find probabilities associated with number resolving detection in distinct spatial modes. The probability for a spatial mode detection pattern described by the vector $\bm{n}=(n_1,\dots,n_{|\mathcal{S}|})$ in spatial modes $\mathcal{S}$ is given by
\begin{equation}
    P_{\mathrm{PNR}}(\mathcal{S};\bm{n}) = \left. \left(\prod_{i=1}^{N}  \frac{(-1)^{n_i}}{n_i!} \partial^{n_i}_{t_i}\right)\det\left(\mathds{1}_{2|S|} + \frac{T_S \tilde\sigma_S T_S}{2} \right)^{-1/2} \right|_{T = \mathds{1}_{2|S|}} \, ,
    \label{eq:numberprob}
\end{equation} 
where $T_S = \mathds{1}_2 \otimes (t \otimes \mathds{1}_{|S|})$ with $t = \diag(t_1, \dots, t_{|\mathcal{S}|})^{1/2}$. 

\end{enumerate}

Eq.~({\ref{eq:numberprob}}) constitutes one of the major results of this work. It allows for the calculation of number resolved detection probabilities across multiple spatial modes, within which multiple spectral degrees of freedom may be present. It should be noted that although our expression in Eq.~({\ref{eq:numberprob}}) appears relatively compact, the presence of the derivatives means that there is in principle an exponentially large number of terms involved in the limit of large photon numbers. Indeed, this can be seen in Eq.~({\ref{PXY}}). This is to be expected, however, as it is known that the calculation of photon number probabilities from Gaussian states is in the {\#}P computational complexity class~\cite{quesada2018gaussian}. 

However, the utility of our expression above in fact lies in the way in which the spectral mode degrees of freedom are included. In our expression the size of the matrix entering the determinant scales only with the number of modes, and is fixed with respect to photon number. This should be compared to the other number resolving detection probability involving matrix Hafnians in Ref.~\cite{quesada2018gaussian}, for which no distinction between spatial and spectral modes is made. As such, we can here include many spectral modes by simply (linearly) increasing the size of the covariance matrix $\sigma_S$, without effecting the scaling of the computation with respect to photon number. To do so using the expressions in Ref.~\cite{quesada2018gaussian}, one would need to calculate matrix Hafnians for all of the different patterns of spectral modes in which photons could have been detected, thus exponentially increasing the number of terms being calculated. Our expressions permit, for example, one to consider arbitrary photon spectra, purity, and indistinguishability.

Finally we note that making the set of modes explicit in the argument in Eq.~({\ref{eq:numberprob}}) clarifies detection patterns when other spectator modes are to be traced out. In order that we have a similar notation for threshold detectors, we introduce a list $\vec{n}$ for threshold detectors which is a click pattern, analogous to the photon number resolving (PNR) click pattern $\vec{n}$ in Eq.~({\ref{eq:numberprob}}), but each element can take on values of only `on' or `off'. We then define
\begin{equation}
\eqalign{
    P_\mathrm{Thres}(\mathcal{S};\vec{n}) = \mathcal{P}_\mathrm{on,off}(\mathcal{X,Y}),
}
\label{eq:P_D}
\end{equation}
where it is understood that the set $\mathcal{X}$ is those modes for which $n_i$ = `on', and $\mathcal{Y}$ those modes for which $n_i$ = `off'. This notation allows us to specify, for either detector type, the set of modes we keep from the whole system $S$ and the particular detector pattern $\vec{n}$. For example we may wish to consider the first four (of potentially greater than four) spatial modes which we label with italicised numbers, $\mathcal{S}=\{\it{1,2,3,4}\}$. With number resolving detectors we then specify the photon numbers in each mode, e.g. $\vec{n}=(1,0,2,0)$, while for threshold detectors we specify the click pattern, e.g. $\vec{n}= ( \mathrm{on}, \mathrm{off}, \mathrm{on}, \mathrm{off})$, and in Eq.~({\ref{eq:P_D}}) we should take $\mathcal{X} = \{1,3\}$ and $\mathcal{Y} = \{2,4\}$. Of course, in general $P_\mathrm{Thres}(\mathcal{S};\vec{n})\neq P_{\mathrm{PNR}}(\mathcal{S};\bm{n})$, so this common notation only indicates a correspondence and not equality.

\subsection{Two-mode squeezers}

Having introduced Gaussian states and photon detection in general terms, we now explore how to describe specific optical elements within this formalism. We begin with the parametric photon pair sources themselves, which arise from Hamiltonians that take the form of multimode two-mode squeezers. These have the general form 
\begin{equation}
\hat{H} = \iint d \nu_1 d \nu_2 F(\nu_1,\nu_2) \hat{a}_{1}^{\dagger}(\nu_1) \hat{a}_{2}^{\dagger}(\nu_2)+\mathrm{h.c.},
\label{HIntegralForm}
\end{equation}
where $\hat{a}_i^{\dagger} (\nu)$ is the creation operator for a mode with spatial mode $i$ and frequency $\nu$. 
Note that at this stage the spectral degree of freedom is treated as a continuous variable. We 
will refer to the function $F(\nu_1,\nu_2)$ as the joint-spectral-amplitude (JSA), the properties of which have a significant effect on the quality of a single photon source based on a two-mode squeezer. 
To see this we write the JSA in its Schmidt decomposition~\cite{law2000continuous,Grice1997}, namely 
\begin{equation}
F(\nu_1,\nu_2) = \sum_l \lambda_l \psi_l (\nu_1) \phi_l(\nu_2)^*,
\label{SchmidtDecomposition}
\end{equation}
where the $\lambda_l$ are positive coefficients known as the Schmidt coefficients, and the functions $\{\psi_l(\nu)\}$ and $\{\phi_l(\nu)\}$ are each sets of orthonormal functions (though not necessarily equal or mutually orthonormal). In the low squeezing limit $\norm{F(\nu_1,\nu_2)} \ll 1 $ the propagator can be expanded to first order to give the bi-photon state
\begin{equation}
    \ket{\Psi} =\e^{-i\hat{H}}\ket{\mathrm{vac}}\approx \ket{\mathrm{vac}}-i \sum_l \lambda_l \hat\mathcal{C}^\dagger_l \hat\mathcal{D}^\dagger_l \ket{\mathrm{vac}},
\end{equation}
where we have introduced the broadband mode operators, $\hat \mathcal{C}^\dagger_l = \int d\nu_1 \psi_l(\nu_1) \hat a^\dagger_1(\nu_1)$ and $\hat \mathcal{D}^\dagger_l = \int d\nu_2 \phi_l^*(\nu_2) \hat a^\dagger_2(\nu_2)$, which themselves are mutually orthogonal. We can therefore see that when the Schmidt decomposition has more than one term the biphoton state is entangled, and when detecting one of the photons in an unknown spectral mode, the other will be left in a spectrally mixed state. Conversely, JSA separability guarantees a pure quantum state after detection of one of the photons. For these reasons we will refer to a source described by a separable JSA as a spectrally pure source.

The continuous nature of the JSA function $F(\nu_1,\nu_2)$ means that the Schmidt basis defined by the functions $\{\psi_\nu(\w)\}$ and $\{\phi_\nu(\w)\}$ must be found by solving integral eigenvalue equations. In practice, it is often easier to instead discretise the integral in Eq.~({\ref{HIntegralForm}}) which, again, is valid under reasonable regularity conditions~\cite{Quesada2019}. In doing so we find the Hamiltonian can be written in precisely the form of Eq.~({\ref{BilinearH}}), where the matrix of coefficients takes the form 
\begin{equation}
\mathbb{H}=\frac{1}{2}\left(\begin{array}{cc}
0 & \mathcal{F} \\
\mathcal{F}^* & 0 \end{array}\right),
\quad \mathrm{with} \qquad
\mathcal{F}=\left(\begin{array}{cc}
0 & F \\
F^\top & 0 \end{array}\right),
\end{equation}
and the matrix $F$ has elements $F_{\w_1\w_2} = \delta \nu F(\delta \nu \w_1,\delta \nu \w_2)$, with $\delta\nu$ being the discretisation step in frequency. The Schmidt decomposition of Eq.~({\ref{SchmidtDecomposition}}) in the discrete case is the singular value decomposition:
\begin{equation}
F= U F_D V^{\dagger},
\label{FSVD}
\end{equation}
where $F_D$ is a diagonal matrix of the singular values of $F$, and $U$ and $V$ are unitary matrices. We note that the factor of $\delta\nu$ in the definition of the matrix elements of $F$ means the singular values in $F_D$ approximate the true singular values $\lambda_l$ in Eq.~({\ref{SchmidtDecomposition}}). 

Eq.~({\ref{FSVD}}) allows the Hamiltonian coefficients to be written $\mathbb{H}=\sfrac{1}{2}\mathbb{U} \mathbb{F}_D \mathbb{U}^{\dagger}$, where 
\begin{equation}
\mathbb{F}_D=\left(\begin{array}{cc}
0 & \mathcal{F}_D \\
\mathcal{F}_D & 0 \end{array}\right),
\quad \mathrm{and} \qquad
\mathbb{U}=\left(\begin{array}{cc}
\mathcal{U} & 0 \\
0 & \mathcal{U}^* \end{array}\right),
\end{equation}
with 
\begin{equation}
\mathcal{F}_D=\left(\begin{array}{cc}
0 & {F}_D \\
{F}_D & 0 \end{array}\right),
\quad \mathrm{and} \qquad
\mathcal{U}=\left(\begin{array}{cc}
{U} & 0 \\
0 & {V}^* \end{array}\right),
\end{equation}
and we note that $\mathcal{U}$ and $\mathbb{U}$ are unitary. With $\mathbb{H}$ written in this way we can perform the exponentiation in Eq.~({\ref{MfromH}}) straightforwardly, and find that the symplectic transformation for the multimode two-mode squeezer can be written $M = \mathbb{U} M_D \mathbb{U}^{\dagger}$ where
\begin{equation}
\label{M2MS}
\hspace{-2cm}
M_D = \left(\begin{array}{cc}
\cosh \mathcal{F}_D & -i \sinh \mathcal{F}_D \\ 
i \sinh \mathcal{F}_D & \cosh \mathcal{F}_D \end{array}\right)
= \left(\begin{array}{cccc}
\cosh {F}_D & 0 & 0 & -i \sinh {F}_D \\ 
0 & \cosh {F}_D & -i \sinh {F}_D & 0 \\ 
0 & i \sinh {F}_D & \cosh {F}_D & 0 \\ 
i \sinh {F}_D & 0 & 0 & \cosh {F}_D \end{array}\right).
\end{equation}
We see that a multimode two-mode squeezer is simply a set of independent two-mode squeezers acting on the appropriate Schmidt modes.

\subsection{Unitary and passive transformations}

In addition to two-mode squeezers which describe parametric sources, we also require unitary mode transformations such as beam-splitters and phase shifters. In terms of symplectic transformations as described in Eq.~({\ref{SymplecticCondition}}), these take the general block-diagonal form $M=\mathrm{diag}(\alpha,\alpha^*)$ with $\alpha^{\dagger}=\alpha^{-1}$.
For a dispersionless (frequency independent) beam-splitter we have 
\begin{equation}
     \alpha_{\mathrm{BS}}(\theta) =
    \left(\begin{array}{cc}
    \cos(\theta) &-\sin(\theta) \\
    \sin(\theta) & \cos(\theta) 
    \end{array} \right) \otimes \mathds{1}_{N_f} \, ,
    \label{alphaBS}
\end{equation}
while a dispersionless phase-shifter acting on say spatial mode ${\it{1}}$ is described by $\alpha_{\mathrm{PS}}(\phi) = \mathrm{diag}(\exp[i \phi],1) \otimes \mathds{1}_{N_f}$. A linear frequency dependent phase shift corresponds to a delay in the time domain. Such a transformation (again acting for illustrative purposes here on mode ${\it{1}}$) is described by $\alpha_{\mathrm{delay}}=\mathrm{diag}(\alpha_\tau,\mathds{1}_{N_f})$ with spectral matrix elements $(\alpha_{\tau})_{\omega_1 \omega_2}=\delta_{\omega_1\omega_2}\exp[i \w_1 \delta\nu\, \tau]$. Here $\omega$ is a discrete (integer) index, and we note that the delay $\tau$ should, in our context, be thought of relative to the bandwidth of photons generated by the two-mode squeezing process, which itself is determined by the typical width of the JSA in Eq.~({\ref{HIntegralForm}}).

As well as these unitary transformations, we will also be interested in more general non-unitary yet passive transformations, and in particular those which correspond to loss or filtering. To implement such transformations we add ancillary loss mode(s), then use a unitary beam-splitter type transformation as in Eq.~({\ref{HIntegralForm}}) acting on the mode of interest and ancillary modes, and then trace out the ancillary modes. This can be done analytically and we do not need to explicitly include the ancillary modes in our calculations. From the condition in Eq.~({\ref{SymplecticCondition}}), we know that a general pure covariance matrix will take the form,
\begin{equation}
    \sigma_S = 2 \left(\begin{array}{cc}
    \beta\beta^{\dagger} & \alpha\beta^\top \\
    (\alpha\beta^\top)^* & (\beta\beta^{\dagger})^*
    \end{array}\right)
    + \mathds{1}_{2|S|} \, .
    \label{eq:covariancemat}
\end{equation}
We imagine that Eq.~({\ref{eq:covariancemat}}) describes modes of interest, and ancillary loss modes are initially in the vacuum state. Coupling into the loss modes is a unitary process, and introducing an ancillary mode for each mode in $S$, the unitary can be written in a general block form 
\begin{equation}
    \mathbb{U}_\mathrm{loss} = 
    \left( \begin{array}{cc}
    \mathbb{U}_{SS} & \mathbb{U}_{SL} \\
    \mathbb{U}_{LS} & \mathbb{U}_{LL} 
    \end{array} \right), \,
\end{equation}
where unitarity is ensured by $\mathbb{U}_{SS}\mathbb{U}^\dagger_{SS} + \mathbb{U}_{SL}\mathbb{U}^\dagger_{SL} = \mathds{1}_{2|S|}$, and we are now working in a basis in which ancillary mode operators are ordered after all mode operators of interest. The action of this unitary on the total covariance matrix is then 
\begin{equation}
\mathbb{U}_\mathrm{loss}
    \left( \begin{array}{cc}
    \sigma_S & 0 \\
    0 & \mathds{1}_{2|S|}
    \end{array} \right)
    \mathbb{U}^\dagger_\mathrm{loss} 
    = 
    \left( \begin{array}{cc}
    \mathbb{U}_{SS} \sigma_S \mathbb{U}^\dagger_{SS} + \mathbb{U}_{SL} \mathds{1}_{2|S|} \mathbb{U}^\dagger_{SL} & \cdots \\
    \vdots & \ddots
    \end{array} \right)   \, , 
\end{equation}
where $\mathbb{U}_{SS}=\mathrm{diag}(\mathcal{U}_{SS},\mathcal{U}_{SS}^*)$ and we have omitted modes which we will shortly trace out. The top left block pertains to our modes of interest, and can be written using $\tilde \sigma_S:= \sigma_S - \mathds{1}_{|2S|}$ as $\mathbb{U}_{SS} (\tilde \sigma_S + \mathds{1}_{2|S|}) \mathbb{U}^\dagger_{SS} + \mathbb{U}_{SL}\mathds{1}_2\mathbb{U}^\dagger_{SL}$. Then using the unitary condition we can rearrange to find 
\begin{equation}
    \sigma_S \to \mathbb{U}_{SS} \tilde{\sigma}_S\mathbb{U}^\dagger_{SS} + \mathds{1}_{2|S|} \, .
\end{equation}
This map, described uniquely by the sub-matrix $\mathcal{U}_{SS}$ of the unitary, is valid for all transformations that can be constructed as a unitary transformation acting on our system of interest and vacuum ancillary modes, followed by a trace over the ancillary modes.

For the purposes of modelling frequency-independent loss on all spatial modes we can take $\mathcal{U}_{SS}=\sqrt{1-\epsilon}\mathds{1}_{|S|}$, with $0\leq \epsilon \leq 1$ the loss parameter. For loss acting on, for example, spatial mode ${\it{1}}$ of $2$, we have $\mathcal{U}_{SS}=\mathrm{diag}(\sqrt{1-\epsilon},1)\otimes \mathds{1}_{N_f}$. Spectral filtering can be included as a frequency-dependent loss with an associated filter function $f(\nu)$ such that $(\mathcal{U}_{SS})_{\omega_1 \omega_2}=\delta_{\omega_1\omega_2} f(\omega_1 \delta\nu )$. For example a bandpass filter with central frequency $\nu_0$ and bandwidth $\Delta\nu_f$ is described by
\begin{equation}
    f(\nu) = \left\{ \begin{array}{cc} 
    1 & \mbox{if } \nu_0 - \Delta\nu_f \leq \nu \leq \nu_0 + \Delta\nu_f \\ 
    0 & \mbox{otherwise } 
    \end{array} \right. \,.
\end{equation}
This provides us with all the optical components necessary to model a HOM interference experiment. 

\section{Heralded Hong--Ou--Mandel interference visibilities}

To begin our investigation of HOM interference visibilities, let us first discuss how such a measurement may be performed. To measure a HOM visibility in a manner closest to the original experiment two signals are interfered on a balanced beam-splitter, and coincidences at the outputs are compared when the signals are made to be as indistinguishable as possible (by, for example, tuning their arrival times to be equal), and when they are made distinguishable by varying some degree of freedom in one of the signals (typically delaying the arrival time of one of the signals).

However, in practice it is not always straightforward to toggle a distinguishability degree of freedom in this way. In particular, typically HOM visibilities measured in integrated silicon platforms are performed by instead scanning through beam-splitter angles using a Mach-Zehnder interferometer~\cite{paesani2020near}, as there is no straightforward way to create a temporal delay with the third order non-linearity in silicon. The maximum and minimum coincidences as a function of beam-splitter angle are then used to derive a visibility. Conversely, other platforms such as Ti:LiNbO$_3$ have a second order nonlinearity and are able to use the polarisation degree of freedom to create a time delay~\cite{luo2019nonlinear}. 
We introduce the term \textit{interference parameter} to allow us to compare different degrees of freedom in one framework. A measured HOM visibility is then potentially dependent on a) the interference parameter used, b) the detector type used, and c) the visibility function used to combine raw counts into a figure of merit. 

\begin{figure}[t!]
\centering
\begin{subfigure}[t]{0.32\textwidth}
    \centering
    \caption{}
    \label{fig:delayhom}
    \resizebox{5cm}{2.1cm}{
    \tikzstyle{block} = [draw,fill=blue!20,minimum size=3em]
    \tikzstyle{smblock} = [draw,fill=blue!20,minimum size=1em]
    \begin{tikzpicture}
        \foreach \y in {0,...,3}
        {
        \pgfmathsetmacro{\addone}{int(\y+1)}
            \node at (0,-0.7*\y) (a\y) {$\textit{\addone}$};
            \node[right of=a\y, node distance = 5.8cm] (det\y) {};
            \draw (a\y) -- (det\y);
            \detector{gray}{5.8}{-0.7 * \y -0.2};
        };
    \node[block] at (1.5,-0.35) (sq) {TMSq$_1$};
    \node[block] at (1.5,-1.75) (sq) {TMSq$_2$};
    \node[smblock] at (3., -0.7) (delay) {$\Delta \tau$};
    \node[block] at (4.5, -1.05) (bs1) {BS($\pi/4$)};
\end{tikzpicture}}
\end{subfigure}
\hfill
\begin{subfigure}[t]{0.32\textwidth}
    \centering
    \caption{}
    \includegraphics[width=1\linewidth]{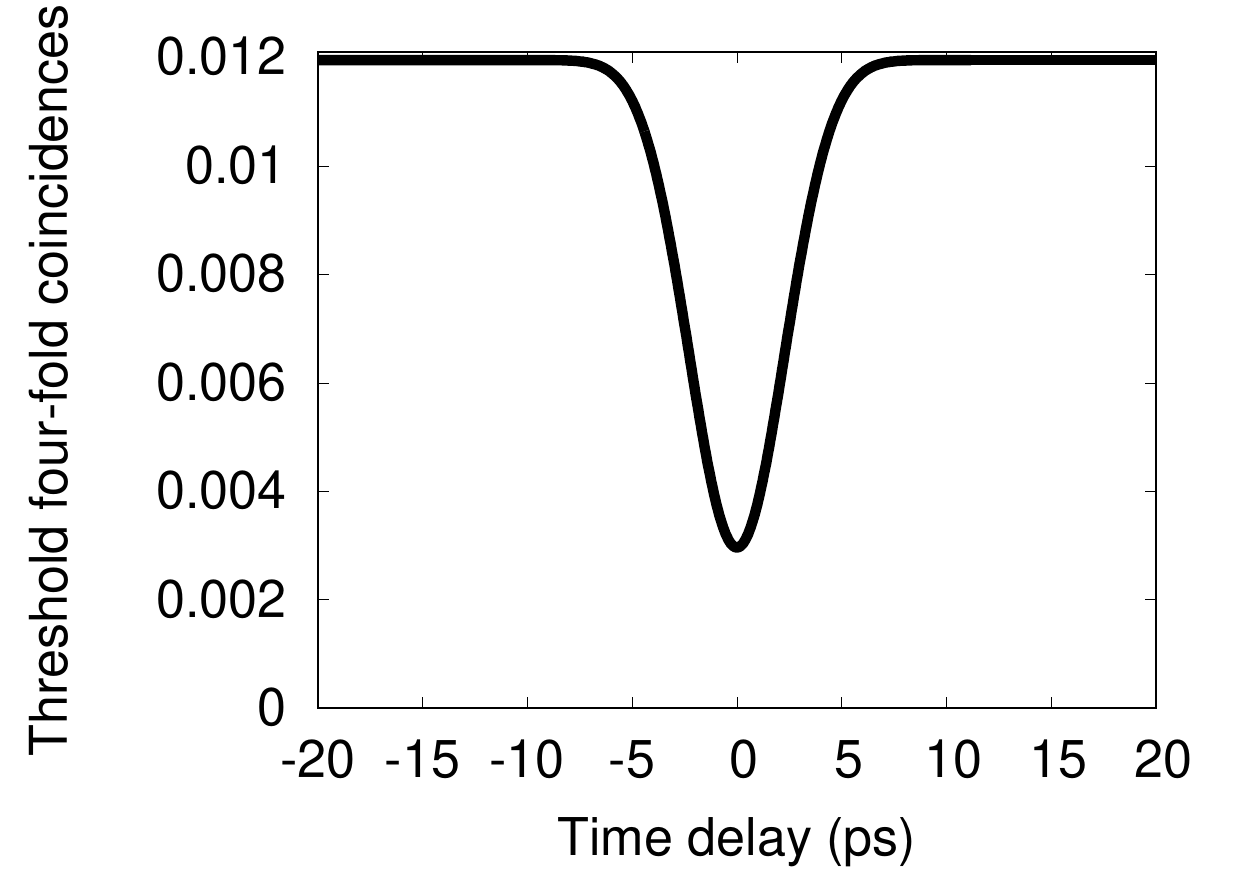}
\end{subfigure}
\hfill
\begin{subfigure}[t]{0.32\textwidth}
    \centering
    \caption{}
    \includegraphics[width=1\linewidth]{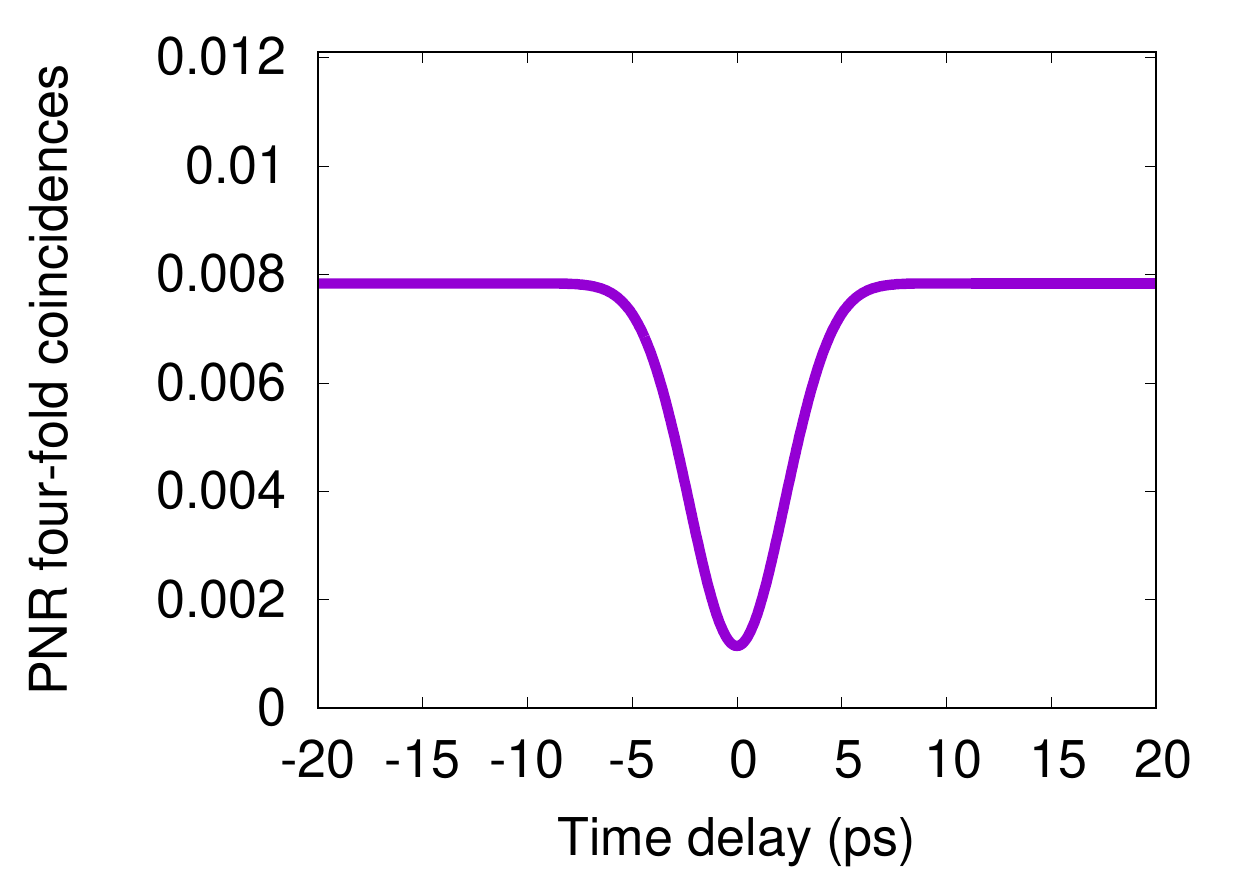}
\end{subfigure}

\begin{subfigure}[t]{0.32\textwidth}
    \centering
    \caption{} 
    \label{fig:mzi}
    \tikzstyle{block} = [draw,fill=blue!20,minimum size=3em]
    \tikzstyle{smblock} = [draw,fill=blue!20,minimum size=1em]
    \resizebox{4.8cm}{2.1cm}{
    \begin{tikzpicture}
        \foreach \y in {0,...,3}
        {
        \pgfmathsetmacro{\addone}{int(\y+1)}
            \node at (0,-0.7*\y) (a\y) {$\textit{\addone}$};
            \node[right of=a\y, node distance = 5cm] (det\y) {}; 
            \draw (a\y) -- (det\y);
            \detector{gray}{5}{-0.7 * \y -0.2};
        };
    \node[block] at (1.5,-0.35) (sq) {TMSq$_1$};
    \node[block] at (1.5,-1.75) (sq) {TMSq$_2$};
    \node[block] at (4.0, -1.05) (bs1) {BS($\theta$)};
\end{tikzpicture}} 
\end{subfigure}
\hfill
\begin{subfigure}[t]{0.32\textwidth}
    \centering
    \caption{} 
    \includegraphics[width=1\linewidth]{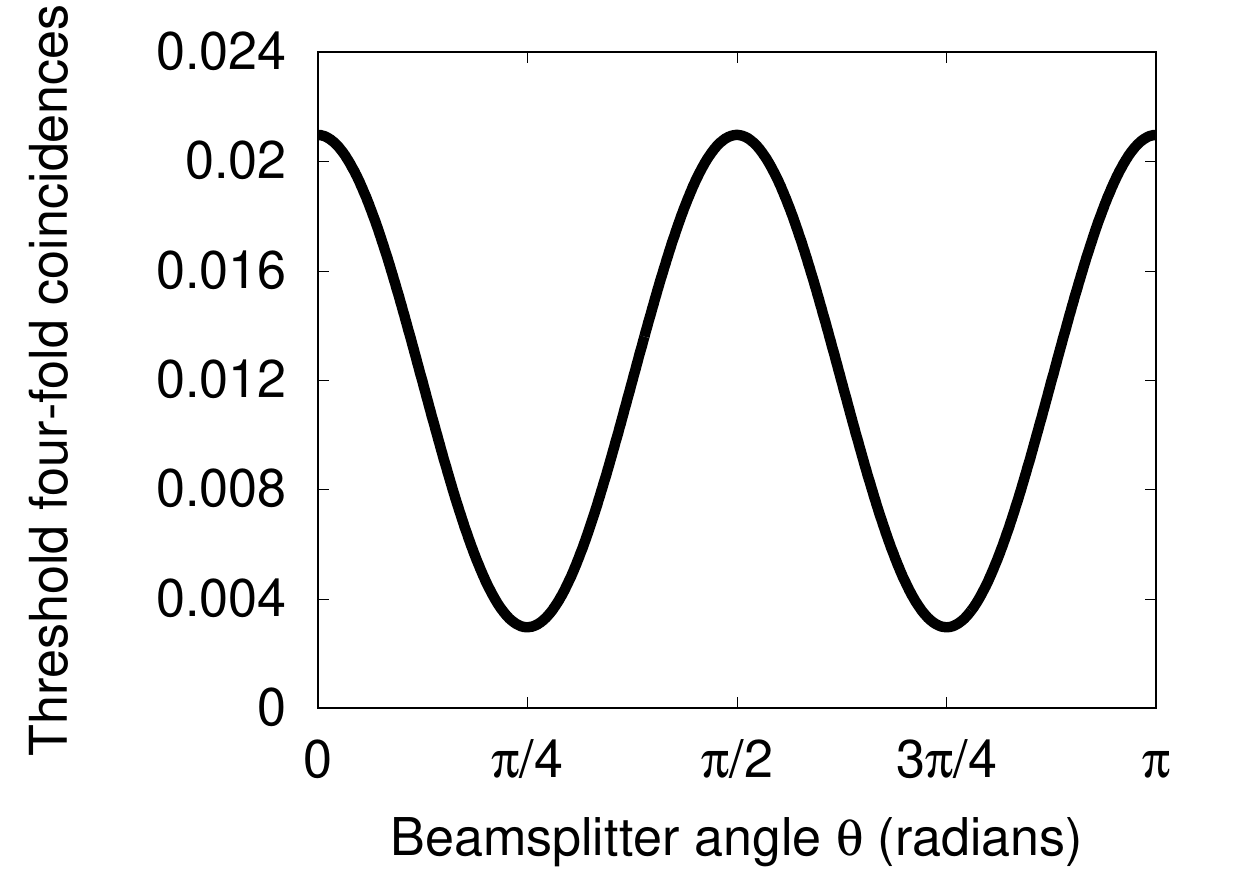}
\end{subfigure}
\hfill
\begin{subfigure}[t]{0.32\textwidth}
    \centering
    \caption{} 
    \includegraphics[width=1\linewidth]{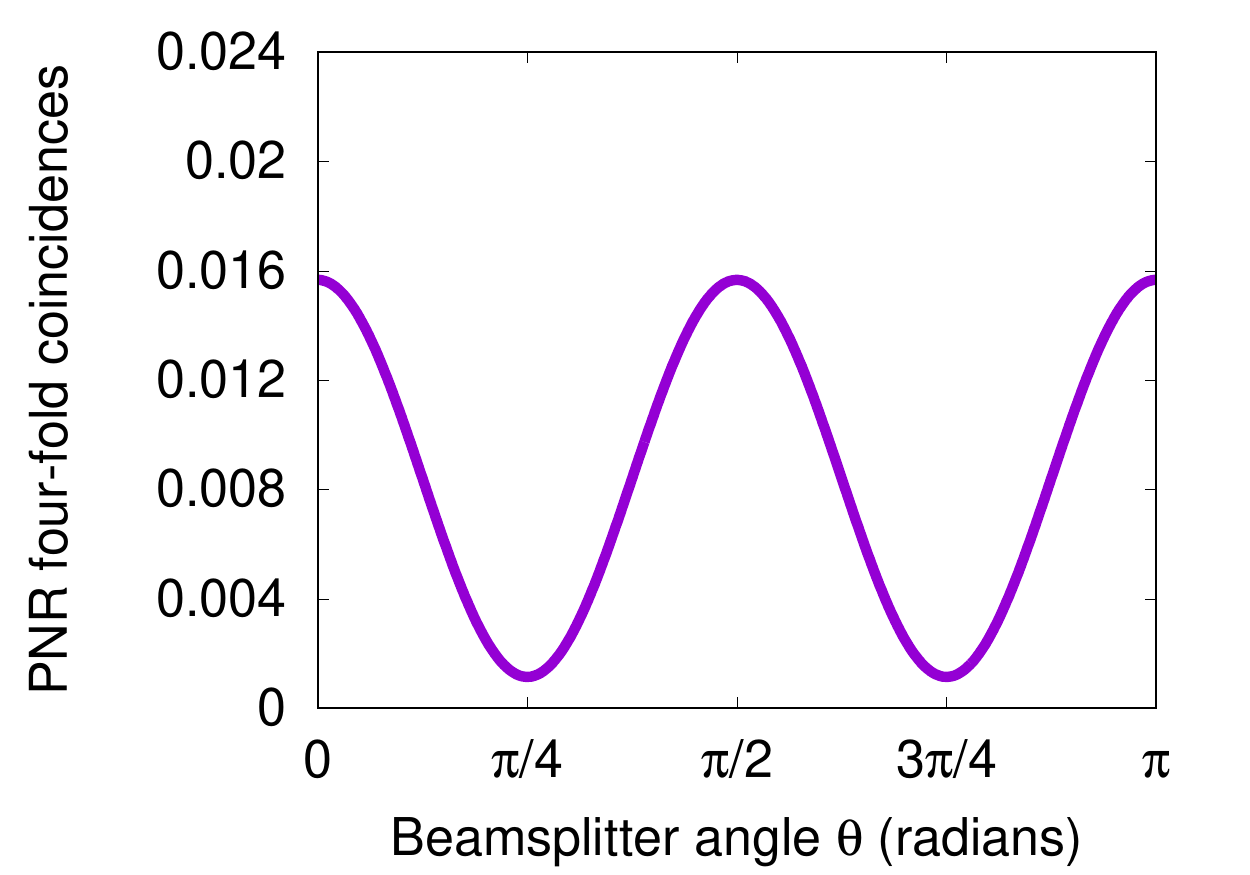}
\end{subfigure}
\caption{Variable time delay (top row) and variable beam-splitter (bottom row) heralded Hong--Ou--Mandel (HOM) measurements. Parts a) and d) show schematics of the optical circuits, while parts b) and e) show the four-fold coincidences as measured using threshold detectors, $P^4_\mathrm{Thres}$ (black), and parts c) and f) using number resolving detectors, $P^4_\mathrm{PNR}$ (purple). Parameters correspond to sources with non-separable JSAs and operating beyond the weak-excitation limit.}
\label{DifferentDipsLab}
\end{figure}
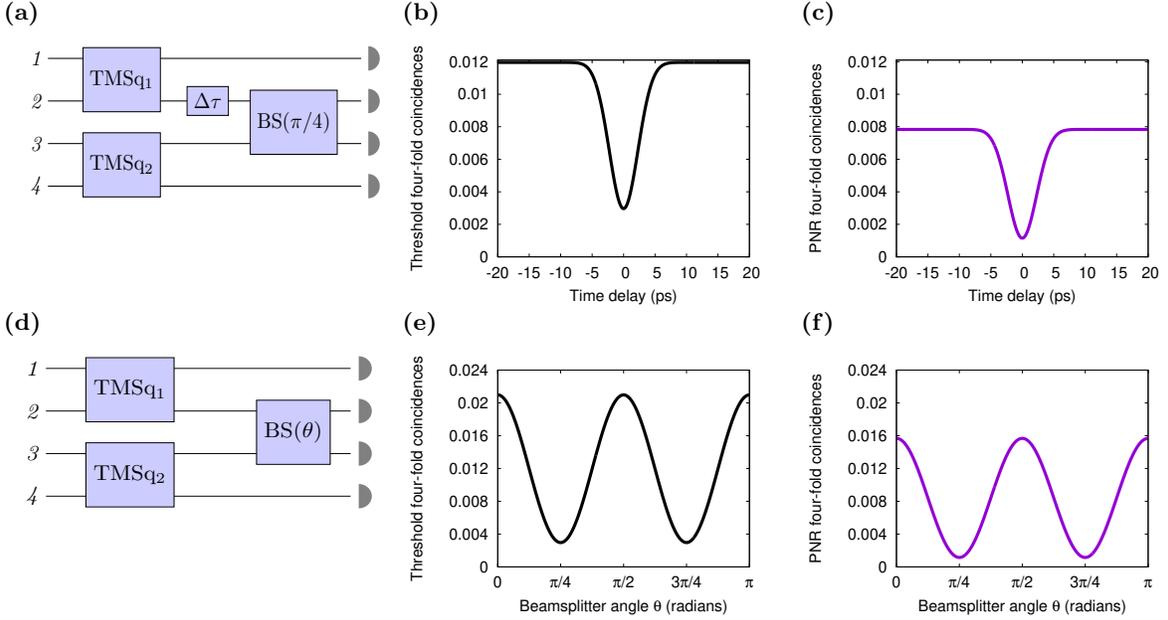

We consider interference between heralded photons from two sources, which we term heralded HOM interference. The optical circuits are shown in Figs.~{\ref{DifferentDipsLab}} a) and d), composed primarily of two sources, a beam-splitter and four detectors. In the ideal case, when the detectors in the two herald {equivalently \emph{signal}} modes ({\it{1}} and {\it{4}}) register the presence of photons, photons are then necessarily present in the two {\emph{idler}} modes ({\it{2}} and {\it{3}}), which then interfere and bunch, leading to a detection event in either mode {\it{2}} or {\it{3}}. We are therefore interested in the four-fold coincidence terms 
\begin{equation}
\eqalign{
    P^\mathrm{4}_\mathrm{PNR} = P_\mathrm{PNR}(\mathit{1,2,3,4}; 1,1,1,1), \\
    P^\mathrm{4}_\mathrm{Thres} = P_\mathrm{Thres}(\mathit{1,2,3,4}; \mathrm{on,on,on,on}),
}
\label{eq:four-fold}
\end{equation}
which should vary from large to small as the interference parameter is increased. For later convenience we will also introduce the bunching terms, which indicate successful heralded HOM interference: 
\begin{equation}
\eqalign{
    P^\mathrm{bunch}_\mathrm{PNR} = P_\mathrm{PNR}(\mathit{1,2,3,4}; 1,0,2,1) + P_\mathrm{PNR}(\mathit{1,2,3,4}; 1,2,0,1), \\
    P^\mathrm{bunch}_\mathrm{Thres} = P_\mathrm{Thres}(\mathit{1,2,3,4}; \mathrm{on,off,on,on}) + P_\mathrm{Thres}(\mathit{1,2,3,4}; \mathrm{on,on,off,on}) \, .
}
\label{eq:bunch}
\end{equation}

In Figs.~{\ref{DifferentDipsLab}} b), c), e), and f) we plot the four-fold coincidences as defined in Eq.~({\ref{eq:four-fold}}) for identical non-separable sources, as a function of time delay (top row) and beam-splitter angle (bottom row). The black curves (parts b) and e)) correspond to coincidences measured with threshold detectors, while the purple curves (parts c) and f)) correspond the same quantity using number resolving detectors. For the varying beam-splitter case, referred as the Mach-Zehnder HOM interferometer, the four-fold coincidence probability has a maximum for a beam-splitter angle of $\theta=0$ (acting trivially on modes {\it{2}} and {\it{3}}) which we label $(P^4_D)_\mathrm{max}$. This is compared to the corresponding value for a balanced beam splitter with $\theta=\pi/4$, which we label $(P^4_D)_\mathrm{min}$. The visibility is typically defined as~\cite{rarity1990two},
\begin{equation}
    V^{\mathrm{MZI}}_D = \frac{(P^4_D)_{\mathrm{max}} - (P^4_D)_{\mathrm{min}}} {(P^4_D)_{\mathrm{max}} + (P^4_D)_{\mathrm{min}}} = \frac{1 - (P^4_D)_\mathrm{min}/(P^4_D)_\mathrm{max}}{1 + (P^4_D)_\mathrm{min}/(P^4_D)_\mathrm{max}}\, .
    \label{eq:vint}
\end{equation}
On the other hand, for a fixed beam-splitter angle of $\theta=\pi/4$ and variable time delay, the coincidences are typically normalised with respect to a large (effectively infinite) delay, which we label $(P^4_D)_\mathrm{\tau=\infty}$. This is then compared to the zero time delay coincidence probability $(P^4_D)_\mathrm{\tau=0}$, which is equal to $(P^4_D)_\mathrm{min}$. We then have the visibility metric~\cite{ou1999photon},
\begin{equation}
    V^{\mathrm{HOM}}_D = 1 - \frac{(P^4_D)_\mathrm{\tau=0}}{(P^4_D)_\mathrm{\tau=\infty}} = 1 - \frac{(P^4_D)_{\mathrm{min}}}{(P^4_D)_\mathrm{\tau=\infty}} \, .
    \label{eq:vhom}
\end{equation}
In both cases the visibility can depend on the type of detector used, and we see that in general these two visibilities are not equal. Even removing the $(P_D^4)_{\mathrm{min}}$ term in the denominator of Eq.~({\ref{eq:vint}}) results in different expressions, as the maximum coincidence probability for variable beam-splitter $(P_D^4)_{\mathrm{max}}$, is not equal to the maximum coincidence probability for variable time delay $(P_D^4)_\mathrm{\tau=\infty}$, leading to a different normalisation in each case. While in the low power limit $(P^{4}_D)_\mathrm{max} = 2(P^{4}_D)_\mathrm{\tau=\infty}$, which reflects the relative number of single photon pathways leading to a coincidence, and gives a single way to equate the two visibilities, this does not hold true away from the low power limit.

\begin{figure}[t!]
\centering
\begin{subfigure}[t]{0.48\textwidth}
    \centering
    \caption{}
    \vspace{-0.4cm}
    \includegraphics[width=1\linewidth]{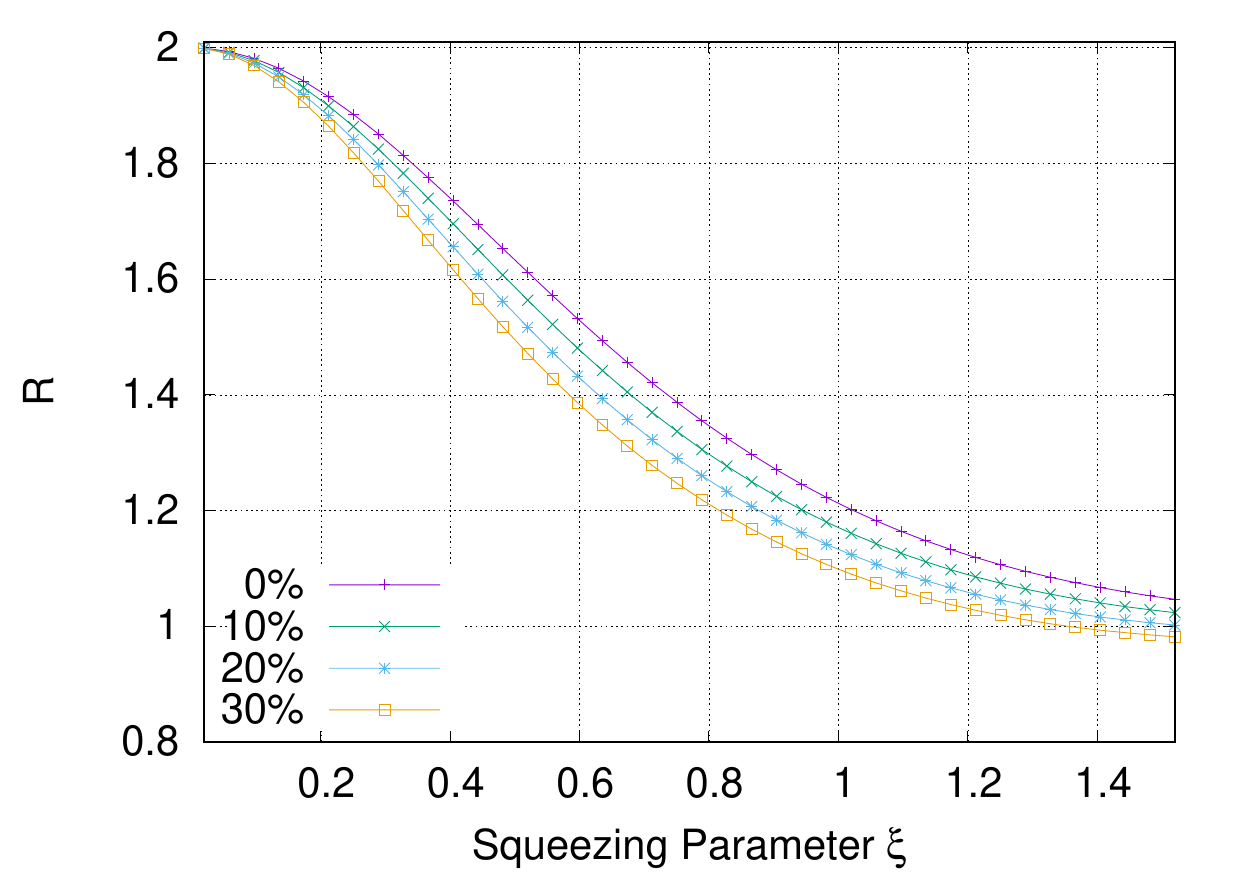}
\end{subfigure}
\hfill
\begin{subfigure}[t]{0.48\textwidth}
    \centering
    \caption{}
    \vspace{-0.4cm}
    \includegraphics[width=1\linewidth]{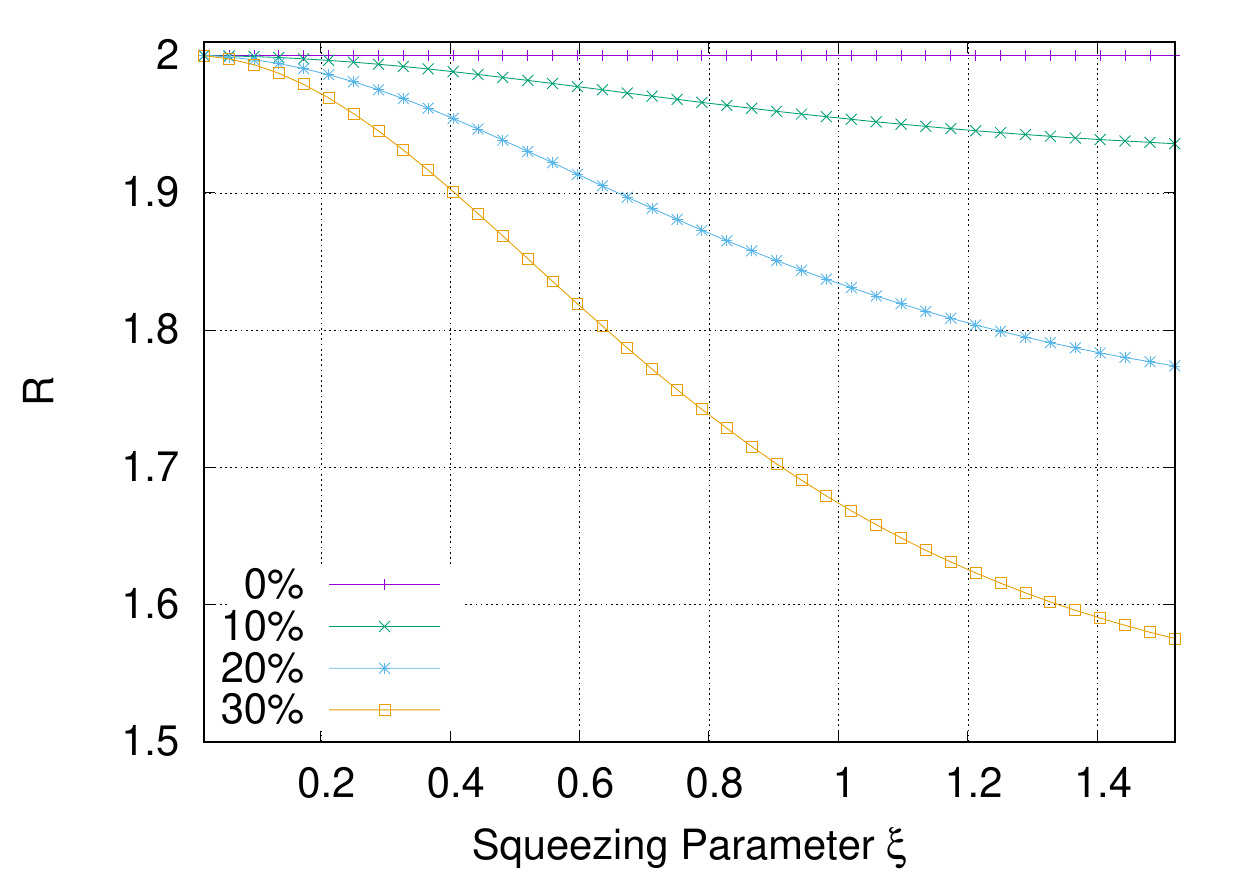}
\end{subfigure}
\caption{The ratio of maximum four-fold coincidences, measured with large time delay $\Delta\tau$ in Fig.~{\ref{DifferentDipsLab}} a), to that measured with zero beam-splitter angle $\theta=0$ in Fig.~{\ref{DifferentDipsLab}} d), i.e.
$(P^4_D)_\mathrm{\tau=\infty} / (P^4_D)_\mathrm{max} = R(\xi, L)$, and for \textbf{(a)} threshold detectors and \textbf{(b)} number resolving detectors, for different loss values as indicated. }
\label{fig:alphanumber}
\end{figure}

To explore this further we can introduce the ratio $R=(P^{4}_D)_\mathrm{max}/(P^{4}_D)_\mathrm{\tau=\infty}$, which allows us to write 
\begin{equation}
    V^\mathrm{HOM}_D = 1 - \frac{ (P^4_D)_\mathrm{min}}{(P^4_D)_\mathrm{\tau=\infty}} = 1 - R\frac{ (P^4_D)_\mathrm{min}}{(P^4_D)_\mathrm{max}} \,.
\label{eq:vnorm}
\end{equation}
We see that with access only to coincidence probabilities measured with variable beam-splitter, knowledge of $R$ allows a visibility calculated in this way to be equated to that measured using a fixed beam-splitter and a time delay. In Fig.~{\ref{fig:alphanumber}} we show how $R$ varies with increasing pump power as captured by the squeezing parameter $\xi$ (defined below), and depending on the type of detector used and level of loss. For threshold detectors we see $R$ decreases with increasing power as coincidence probabilities begin to saturate. However $R$ also decreases for number resolving detectors beyond the low squeezing regime and for non-separable sources. 

These findings demonstrate that some care must be taken in deducing a HOM interference visibility when using any given experimental setup. Although it would seem from Fig.~{\ref{fig:alphanumber}} that either interference parameter can be used provided the measurements are taken in the low power limit, as we will see in the remainder of this paper, interference visibilities are not in general constant with power, even when using number resolving detectors. Moreover, when considering larger scale systems involving more photons, it is unlikely that their successful operation or fidelity with target states will be linear functions of, or even uniquely defined in terms of, these simple HOM interference visibilities. We will see in what follows, however, that in the limiting case of two identical sources and in the absence of loss, it is the visibility $V^{\mathrm{HOM}}_{\mathrm{PNR}}$ measured with variable time-delay and number resolving detectors which gives a direct measure of the heralded photon purity. As such, for the sake of concreteness, we will use use Eq.~({\ref{eq:vhom}}) as our visibility figure of merit for the remainder of this paper, but note that careful consideration of how this figure or merit affects a specific application will be required.

\subsection{Effects of spectral and number impurity}

We now investigate how the pump power simultaneously affects the heralded HOM interference visibility and heralding rate. We first consider sources which have separable JSAs, which give rise to heralded photons that are spectrally pure. To do so we take as an example an idealised JSA that takes the functional form of the product of two Gaussian functions~\cite{paesani2020near},
\begin{equation}
\hspace{-1cm}
F_{\mathrm{Gauss}}(\nu_1,\nu_2) = \frac{\xi}{\norm{F_\mathrm{Gauss}}_\mathrm{Fr}}
\exp\Big[-\frac{1}{2}\left(\frac{\Delta\nu_1}{\zeta}\right)^2\Big] \exp\Big[-\frac{1}{2}\left(\frac{\Delta\nu_2}{\zeta}\Big)^2\right] \, ,
\label{eq:gaussjsa}
\end{equation}
where $\Delta\nu_i = \nu_i - \overline{\nu}_i$ with $\overline{\nu}_i$ are the central frequencies of the signal and idler photons and $\zeta$ represents their bandwidth. The denominator here represents the Frobenius norm of the exponential factor. In this way the JSA features a single Schmidt coefficient given by $\xi$.

\begin{figure}[t!]
\centering
\begin{minipage}[c]{0.33\textwidth}
\centering
\vspace{-0.2cm}
\begin{tikzpicture}
\node[label={[shift={(-2.2,-0.6)}]\textbf{(a)}}] at (0,0) {\includegraphics[width=\textwidth]{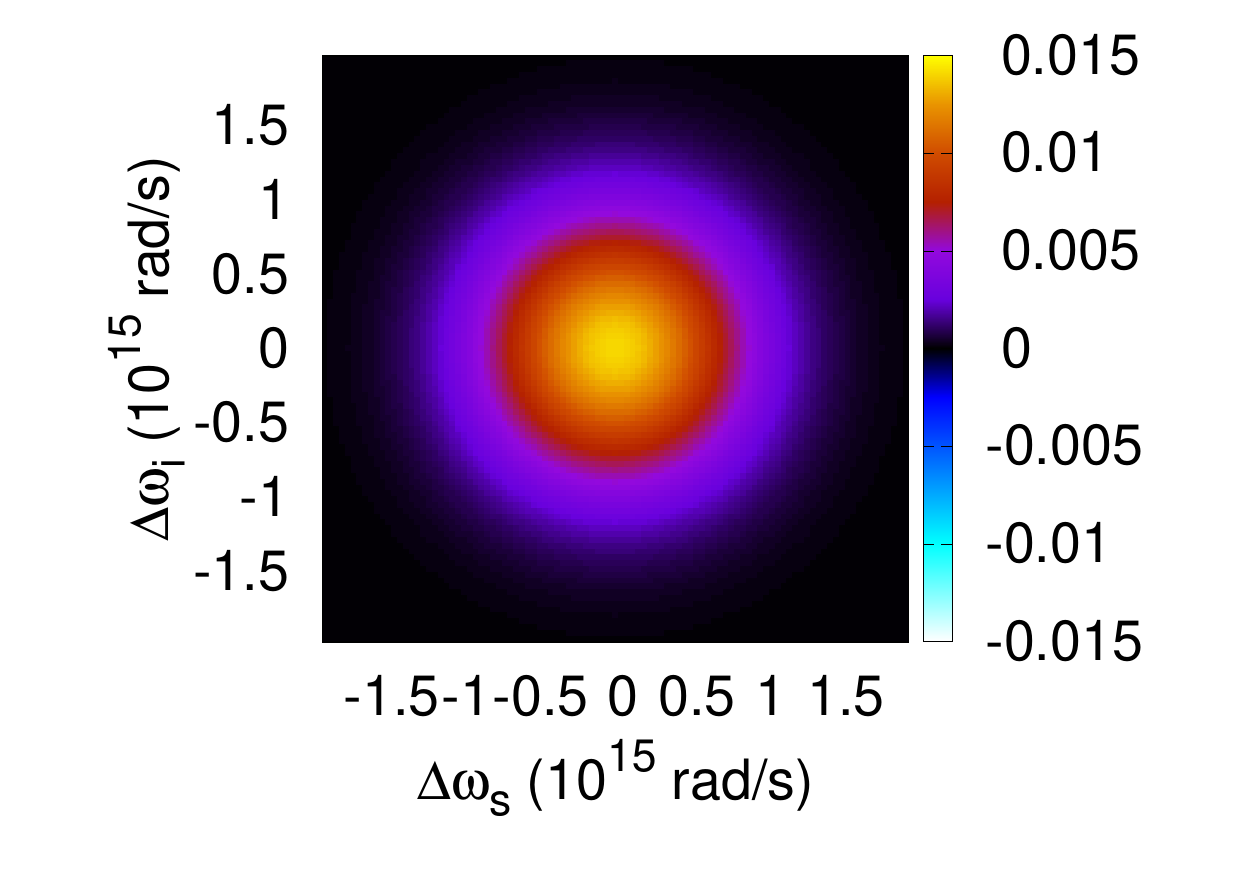}};
\end{tikzpicture}
\vspace{-0.4cm}
\vspace{-0.4cm}
\centering
\begin{tikzpicture}
\node[label={[shift={(-2.2,-0.6)}]\textbf{(b)}}] at (0,0) {\includegraphics[width=\textwidth]{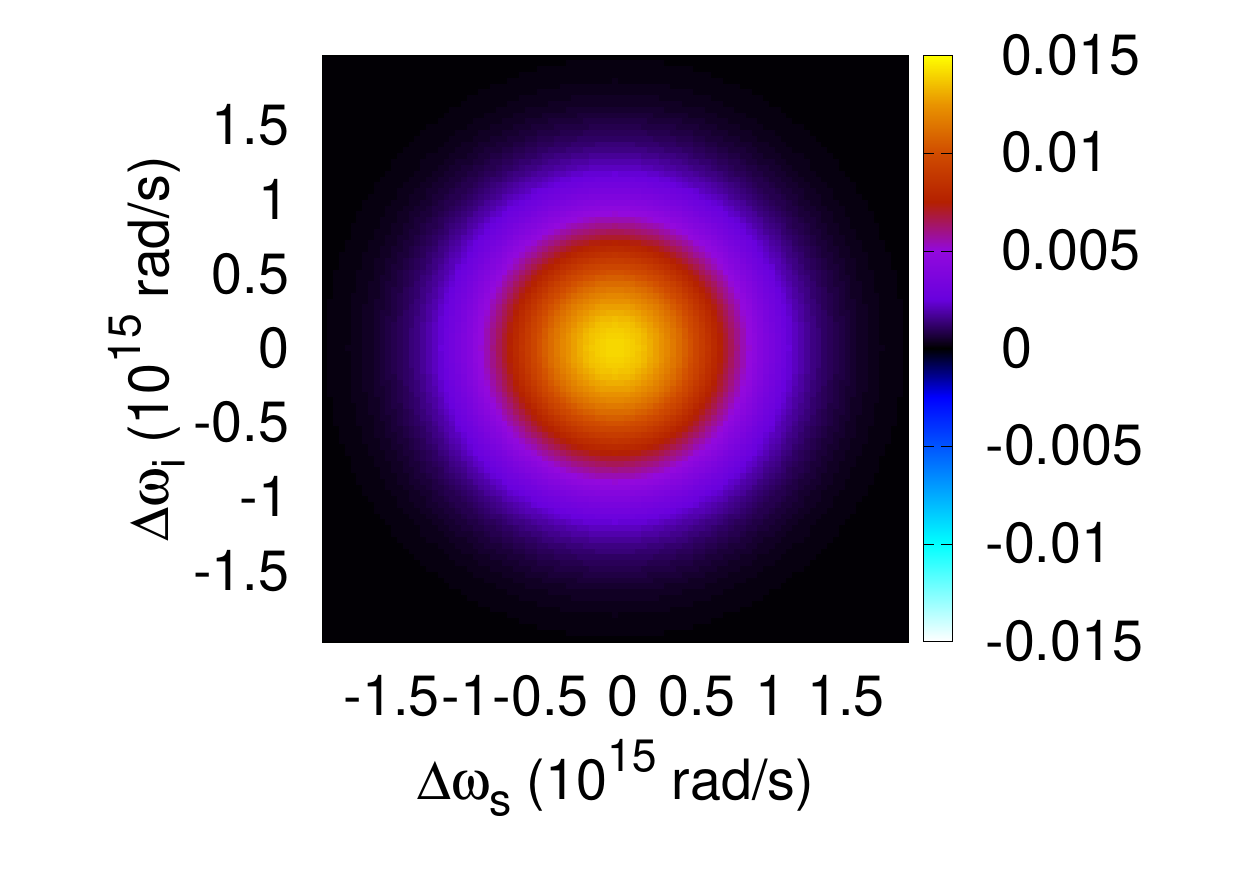}};
\end{tikzpicture}
\end{minipage}
%
%
\begin{minipage}[c]{0.635\textwidth}
\vspace{-0.2cm}
\centering
\begin{tikzpicture}
\node[label={[shift={(-4.4,-0.6)}]\textbf{(c)}}] at (0,0) {\includegraphics[width=\textwidth]{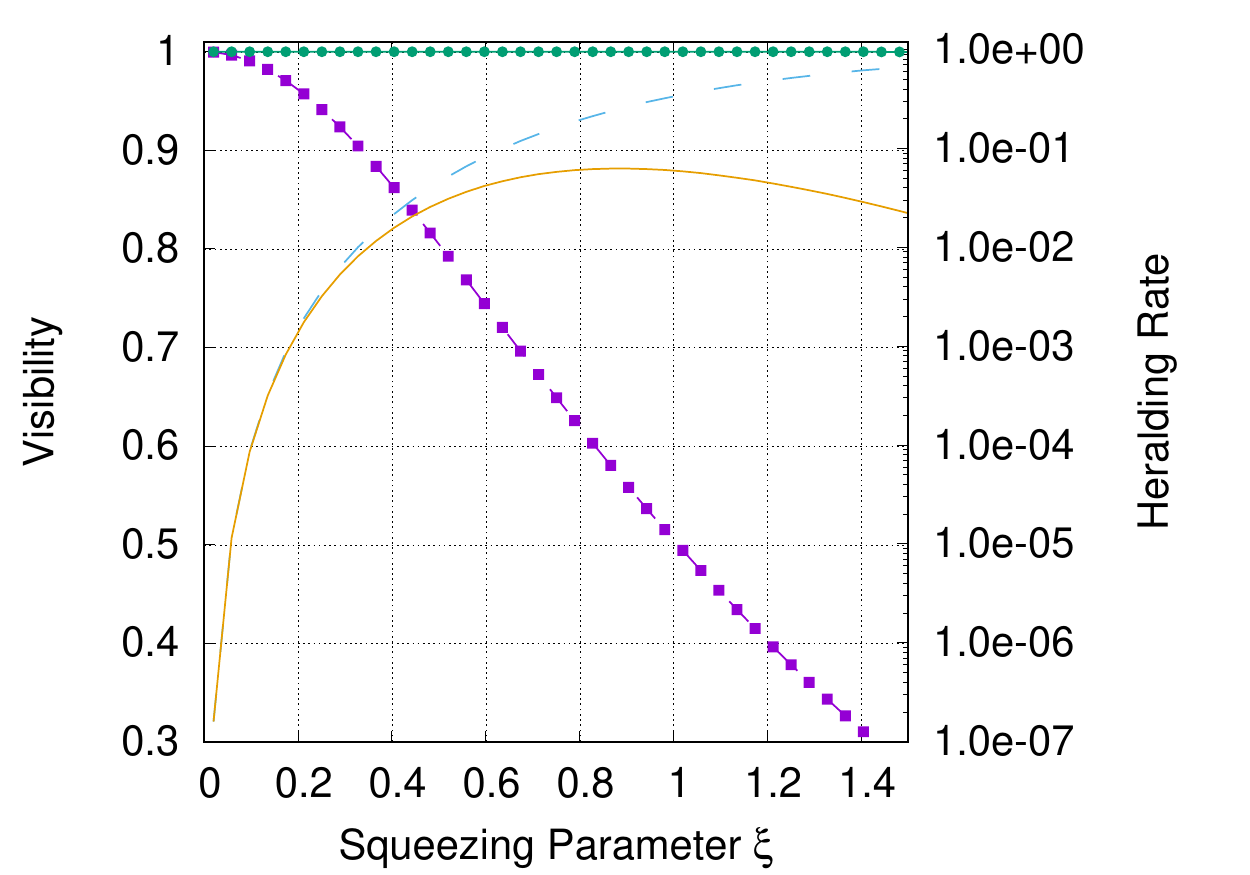}};
\end{tikzpicture}
\label{fig:purehomvisandrate}
\end{minipage}
\caption{Figures of merit for interference between two sources with separable joint-spectral-amplitudes (JSAs). Parts a) and b) show the real part of the source JSAs (which are here equal), part c) shows the joint heralded HOM visibility and heralding rate as a function of squeezing parameter $\xi$ (i.e. pump power). We show the visibility using number resolving detectors $V^\mathrm{HOM}_\mathrm{PNR}$ (green circles) and threshold detectors, $V^\mathrm{HOM}_\mathrm{Thres}$ (purple squares), together with the joint heralding rate for number resolving  detectors $P^\mathrm{Herald}_\mathrm{PNR}$ (yellow solid curve) and threshold detectors $P^\mathrm{Herald}_\mathrm{Thres}$ (blue dashed curve). We use the parameters $\zeta = 0.1$ THz.}
\label{fig:pure}
\end{figure}%

In Fig.~{\ref{fig:pure}} we plot the real part of the JSA of both sources on the left, and on the right we show the heralded HOM visibility and joint heralding rate as a function of the squeezing parameter $\xi$ which represents the pump power. Visibilities are calculated using both threshold (purple squares) and number resolving detectors (green circles). The heralding rate is the probability that both heralding (signal) arms (modes {\it{1}} and {\it{4}}) register photons, and can be thought of as the square of efficiency of one of the sources per excitation pulse. It is defined as 
\begin{equation}
    P^\mathrm{Herald}_{\mathrm{PNR}} = P_{\mathrm{PNR}}(\mathit{1,4};1,1),
    \qquad\mathrm{and}\qquad
    P^\mathrm{Herald}_{\mathrm{Thres}} = P_{\mathrm{Thres}}(\mathit{1,4};\mathrm{on},\mathrm{on})
\end{equation}
for number resolving detectors (yellow curve) and threshold detectors (blue dashed curve), respectively.
We see that for pure sources the number resolved visibility is $1$ for all values of the squeezing parameter. This reflects the fact that regardless of the power, a heralding event selects single photon states in the idler modes, which then perfectly interfere to give no coincidences. The heralding rate, however, begins to decrease for large squeezing parameters as the probability of only single photon events decreases, and the optimal squeezing parameter is approximately $\xi=0.9$, which corresponds to 7.8dB of squeezing. Using threshold detectors, we see that the HOM visibility decreases with increasing squeezing parameter, which is due to the higher order multiphoton terms contaminating the `single photon' state. The heralding rate monotonically increases as it does not distinguish between the single photon and multiphoton subspaces, and eventually saturates at $1$. As we might expect, threshold detectors do reasonably approximate the number resolving heralding rate for squeezing parameter $\xi < 0.2$ (squeezing of 1.7dB). Since the sources here produce spectrally pure photons, Fig.~({\ref{fig:pure}}) serves only to highlight the effects of photon number purity, which we see can be mitigated with the use of number resolving detectors.

We now also include the effects of spectral impurity. To do so we consider two typical waveguide sources which gives rise to non-separable JSAs. The functional form we use is 
\begin{equation}
\hspace{-1cm}
    F_{\mathrm{w.g.}}(\nu_1,\nu_2) = \frac{\xi}{\norm{F_\mathrm{w.g.}}_\mathrm{Fr}} \exp\Big[-\frac{1}{2}\frac{(\Delta\nu_1 + \Delta\nu_2)^2}{\zeta^2}\Big] \mathrm{sinc}\Big[\frac{\tilde L}{2} (\Delta\nu_1  - \Delta\nu_2)\Big] \, ,
    \label{spiralJSA}
\end{equation}
where here the parameter $\tilde L$ captures the fields' propagation over the effective length of the two-mode squeezing interaction. Such a JSA is achieved for spontaneous parametric down-conversion using a $\chi^{(2)}$ non-linearity where $\zeta$ represents the pump bandwidth, or using degenerately pumped spontaneous four-wave mixing under a $\chi^{(3)}$ non-linearity with $\zeta$ being the autoconvolution of the pump. The functional form above is a consequence of symmetric phase-matching with the central frequencies obeying $\bar{\nu}_s+\bar{\nu}_i-(n-1)\bar{\nu}_p= 0$, the central wave-vectors satisfy $\bar{k}_s+\bar{k}_i-(n-1)\bar{k}_p+2\pi/P=0$, for possible poling period $P$, and the fields' group velocities obey $v_{i}^{-1}=2v_{p}^{-1}-v_{s}^{-1}$ so that 
$\tilde L = L (v_{s}^{-1}-v_{p}^{-1})$ with $L$ the interaction length. In all cases the subscripts indicate the signal, idler and pump, and $n=2,3$ is the order of the non-linearity. The normalisation factor in Fig.~({\ref{spiralJSA}}) ensures the JSA has Schmidt coefficients of the form $\lambda_l=\xi\alpha_l$ with $\sum_l \alpha_l^2=1$, and the multiplying factor $\xi$ is the (multi-mode) squeezing parameter which is directly related to the strength of the pump laser. 
\begin{figure}[t!]
\centering
\begin{minipage}[c]{0.33\textwidth}
\centering
\begin{tikzpicture}
\node[label={[shift={(-2.2,-0.6)}]\textbf{(a)}}] at (0,0) {\includegraphics[width=\textwidth]{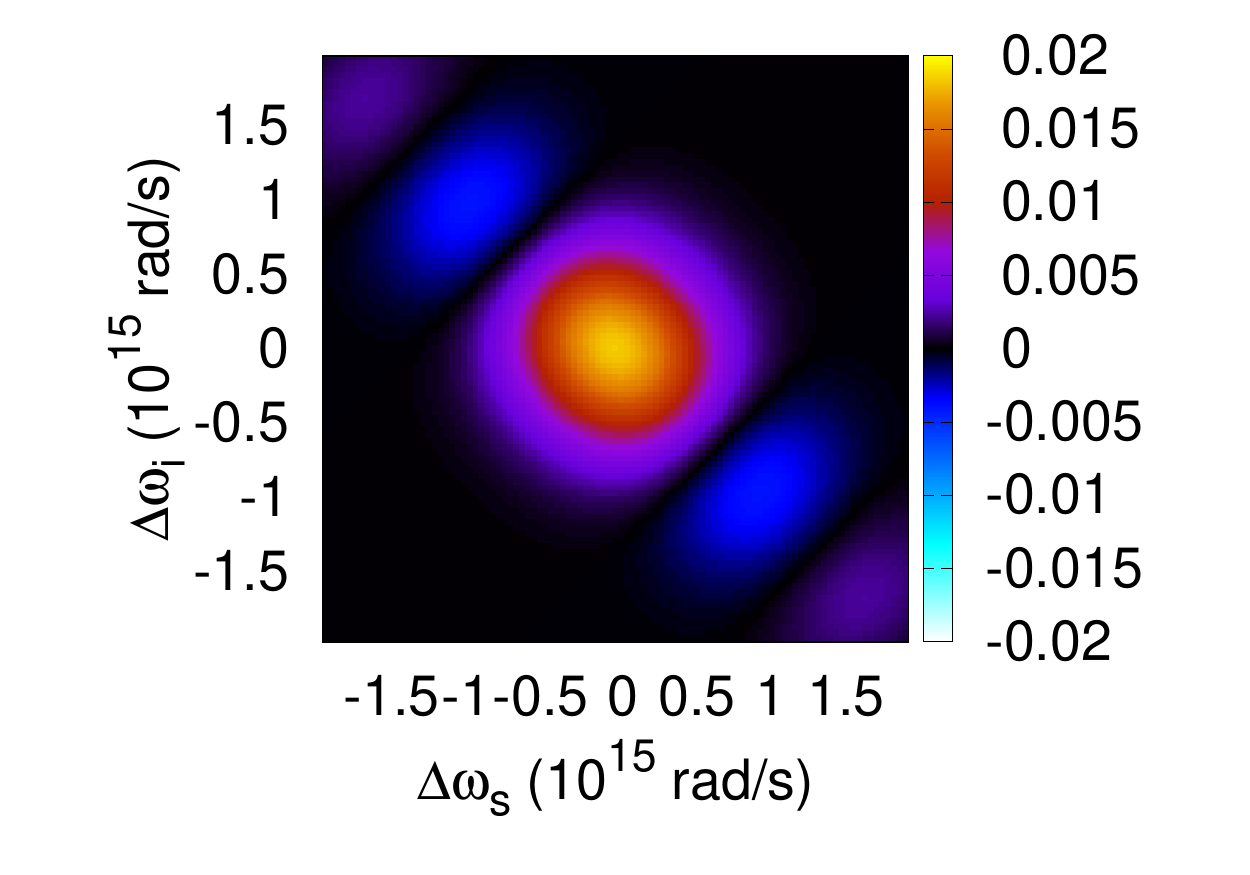}};
\end{tikzpicture}
\vspace{-0.4cm}
\vspace{-0.4cm}
\centering
\begin{tikzpicture}
\node[label={[shift={(-2.2,-0.6)}]\textbf{(b)}}] at (0,0) {\includegraphics[width=\textwidth]{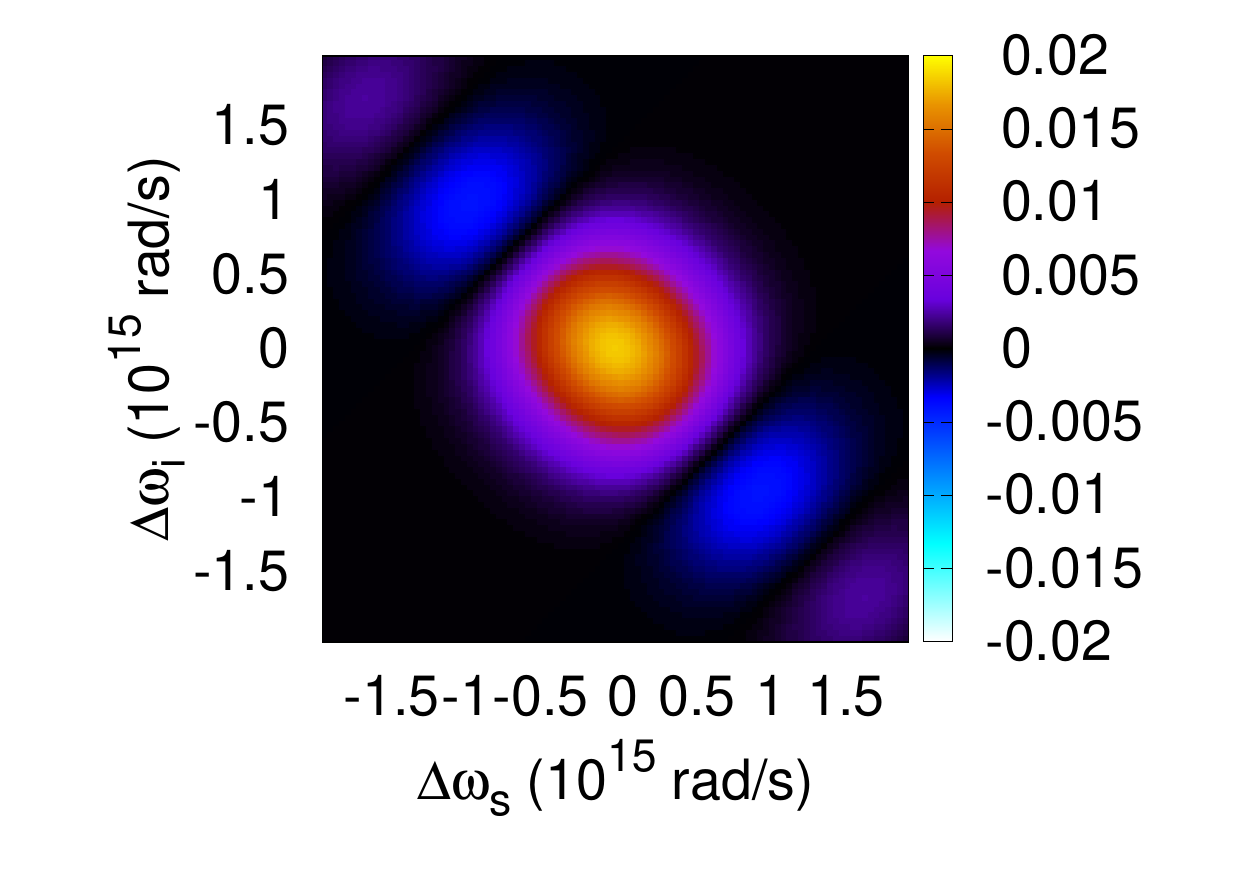}};
\end{tikzpicture}
\end{minipage}
%
%
\begin{minipage}[c]{0.635\textwidth}
\vspace{-0.4cm}
\centering
\begin{tikzpicture}
\node[label={[shift={(-4.4,-0.6)}]\textbf{(c)}}] at (0,0) {\includegraphics[width=\textwidth]{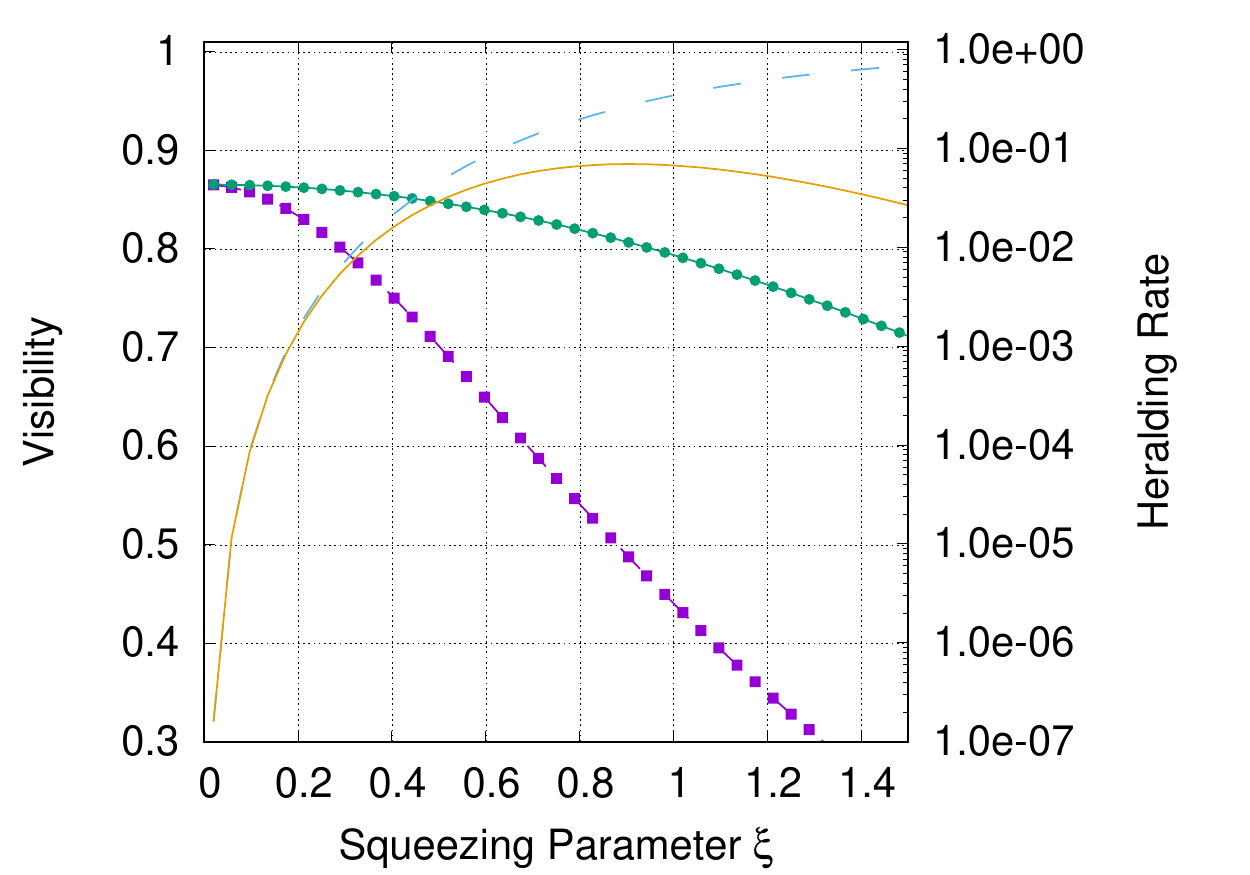}};
\end{tikzpicture}
\end{minipage}
\caption{Figures of merit corresponding to two sources with non-separable (JSAs). Parts a) and b) show the real part of the JSAs (here being equal), while c) is the HOM visibility for number resolving detectors, $V^\mathrm{HOM}_\mathrm{PNR}$ (green circles) which matches exactly with Eq.~({\ref{eq:specpurity}}), and threshold detectors $V^\mathrm{HOM}_\mathrm{Thres}$ (purple squares), together with the joint heralding rate for number resolving detectors $P^\mathrm{Herald}_\mathrm{PNR}$ (yellow curve) and threshold detectors $P^\mathrm{Herald}_\mathrm{Thres}$ (blue dashed curve). We use the parameters $\zeta = 0.1$~THz, $\tilde L = 29.0$~ps.}
\label{fig:sinchomvis}
\end{figure}

In Fig.~({\ref{fig:sinchomvis}}) we again show the real part of the JSA of each source on the left, and the HOM visibility and heralding rate on the right. For threshold detectors, we see similar trends with increasing squeezing as for the pure sources case, though now with the HOM visibility starting below unity, reflecting the fact that even in the single photon subspace interference is imperfect owing to the spectral impurity of the heralded photons. In contrast to the case explored above however, when using number resolving detectors we see that the HOM interference visibility decreases with increasing squeezing parameter~\cite{Christ2012}. This suggests that even when using number resolving detectors to herald photons only in the single photon subspace, the power, as captured by the squeezing parameter, cannot be increased without detrimentally affecting the interference probability of the photons produced. 



We can see this decreasing of the interference visibility by following Ref.~\cite{Christ2012} and using the Schmidt decomposition in Eq.~({\ref{SchmidtDecomposition}}), as it allows the two-mode squeezing Hamiltonian in Eq.~({\ref{HIntegralForm}}) to be written $\hat{H}=\sum_l \lambda_l \hat{\mathcal{C}}_{l}^{\dagger} \hat{\mathcal{D}}_{l}^{\dagger}+\mathrm{h.c.}$, where $\hat{\mathcal{C}}_{l}^{\dagger}=\int d \nu \psi_l(\nu) \hat{a}_1^{\dagger}(\nu)$ and $\hat{\mathcal{D}}_{l}^{\dagger}=\int d \nu \phi_l^*(\nu) \hat{a}_2^{\dagger}(\nu)$ are the generalised Schmidt modes as before. Since these generalised modes are independent, the unitary time evolution operator corresponding to a multi-mode two-mode squeezing operation is in fact just a product of independent two-mode squeezers, and we have 
\begin{equation}
\hat{U}=\exp[-i\hat{H}]=\bigotimes_l \hat{S}_{l}^{(2)}(-i\lambda_l),
\end{equation}
where 
$\hat{S}_{l}^{(2)}(z)=\exp[z\hat{\mathcal{C}}_{l}^{\dagger}\hat{\mathcal{D}}_{l}^{\dagger}-z^*\hat{\mathcal{C}}_{l}\hat{\mathcal{D}}_{l}]$ is a two-mode squeezer acting on spectral mode $l$ in spatial modes ${\it{1}}$ and ${\it{2}}$, and which in normally ordered form is written~\cite{Hong-Yi1987}
\begin{equation}
\hspace{-2cm}
\eqalign{
\hat{S}_{l}^{(2)}(z)
=\exp[\e^{i\phi}\tanh r \, \hat{\mathcal{C}}_{l}^{\dagger}\hat{\mathcal{D}}_{l}^{\dagger}]
\exp[\ln\sech r \, (\mathds{1}+\hat{\mathcal{C}}_{l}^{\dagger}\hat{\mathcal{C}}_{l} + \hat{\mathcal{D}}_{l}^{\dagger}\hat{\mathcal{D}}_{l}]
\exp[-\e^{-i\phi}\tanh r \, \hat{\mathcal{C}}_{l}\hat{\mathcal{D}}_{l}],
}
\end{equation}
where we have written $z=r\e^{i\phi}$. To find the joint signal--idler state produced by a source including all spectral modes and photon numbers, we can act this on the vacuum to give 
\begin{equation}
\ket{\Psi}=\hat{U}\ket{\mathrm{vac}}=\sum_{\vec{n}} c(\vec{n}) \ket{\vec{n}}_1\ket{\vec{n}}_2,
\end{equation}
where the sum runs over all possible integer tuples indicating the number of photons in each spectral Schmidt mode, i.e. $\vec{n}=(n_{l_1},n_{l_2},\dots)$, while 
$\ket{\vec{n}}_i=\bigotimes_l \ket{n_l}_{i l}$ with 
$\ket{n_l}_{1 l}=\hat{\mathcal{C}}_l^{\dagger n_l} \ket{\mathrm{vac}}_{1 l}/\sqrt{n_l!}$ (and similarly for $\ket{n_l}_{2 l}$) represents the corresponding Fock state in spatial mode $i$,
and the coefficients are $c(\vec{n})=\prod_l \sech \lambda_l (-i \tanh \lambda_l)^{n_{l}}$. 

To find the interference probability of the idler photons, we consider the conditional state obtained when a single photon in any Schmidt mode is detected in spatial mode ${\it{1}}$, i.e. the signal mode. The corresponding measurement operator is the projector $\Pi_1 = \sum_{l} \ketbra{1}{1}_{1 l}$. The probability
for such a detection is $P_{\mathrm{PNR}}({\it{1}};1)=\mathrm{Tr}[\ketbra{\Psi}{\Psi}\Pi_1]$, giving 
\begin{equation}
P_{\mathrm{PNR}}({\it{1}};1)=\Big(\prod_l\sech\lambda_l\Big)^2 \sum_l \tanh^2\lambda_l,
\end{equation}
while the post measurement state is $\hat{\rho}=\mathrm{Tr}_1[\ketbra{\Psi}{\Psi}\Pi_1]/P_{\mathrm{PNR}}({\it{1}};1)$ with the trace only over modes with spatial label ${\it{1}}$, which gives 
\begin{equation}
\hat{\rho}=\Big(\sum_l \tanh^2\lambda_l\Big)^{-1}\sum_l \tanh^2\lambda_l \ketbra{1}{1}_{2l}.
\end{equation}
The purity of this single photon state is a measure of its spectral indistinguishability. For this we find 
\begin{equation}
\mathrm{Tr}[\hat{\rho}^2]=\Big(\sum_l \tanh^2\lambda_l\Big)^{-2}\sum_l \tanh^4\lambda_l,
\label{eq:specpurity}
\end{equation}
which we see depends on the distribution of the Schmidt coefficients. If one Schmidt mode dominates and is much larger than all others, we have simply $\mathrm{Tr}[\hat{\rho}^2]\approx 1$. However, if this is not the case, then we see that linearly increasing each coefficient causes a decrease in the idler photon purity. Eq.~({\ref{eq:specpurity}}) is plotted with black curve in Fig.~{\ref{fig:sinchomvis}}, and perfectly matches the green circular data points as expected.

\subsection{Effects of photon loss and filtering}

\begin{figure}[t!]
\centering
\begin{subfigure}[c]{0.48\textwidth}
\centering
\caption{}
\vspace{-0.4cm}
\includegraphics[width=1\textwidth]{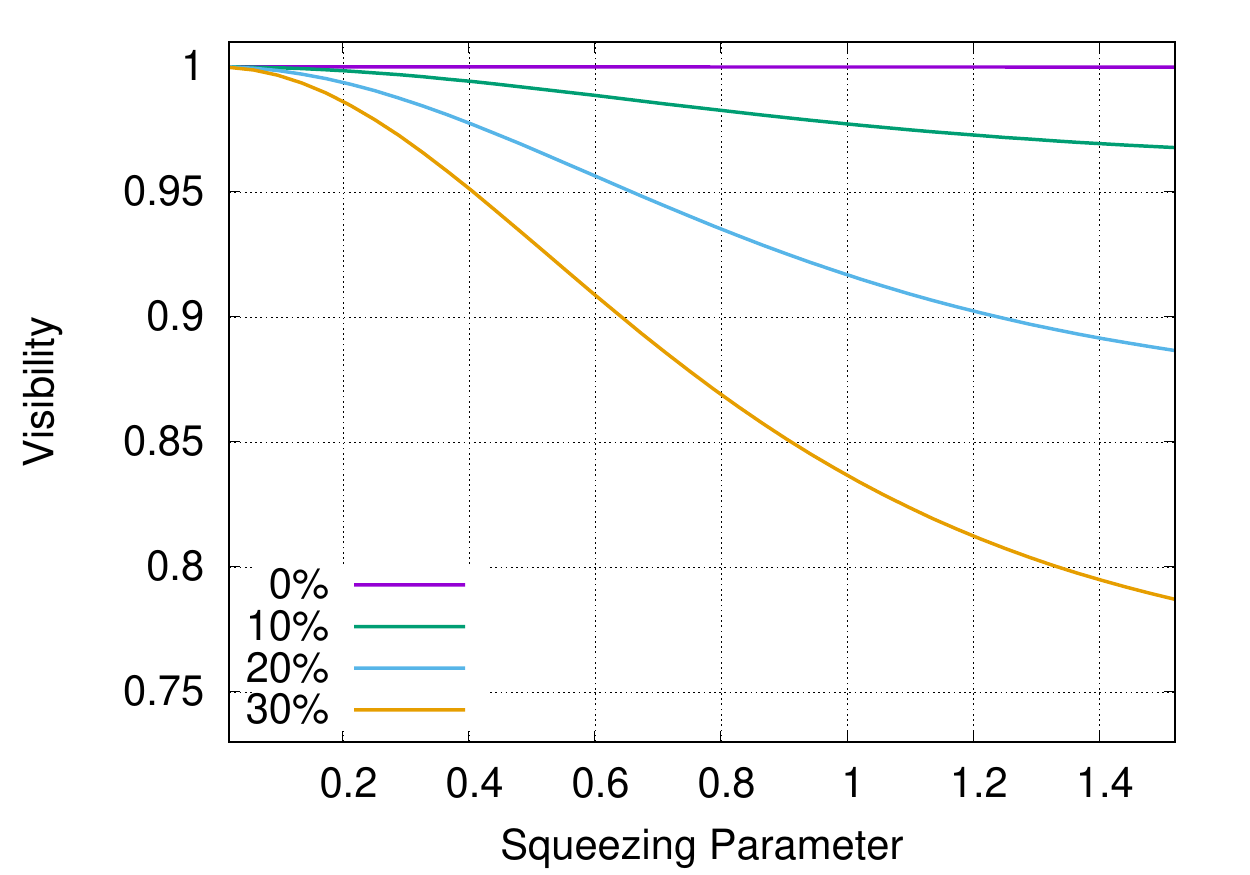}
\end{subfigure}
\hfill
\begin{subfigure}[c]{0.48\textwidth}
\centering
\caption{}
\vspace{-0.4cm}
\includegraphics[width=1\textwidth]{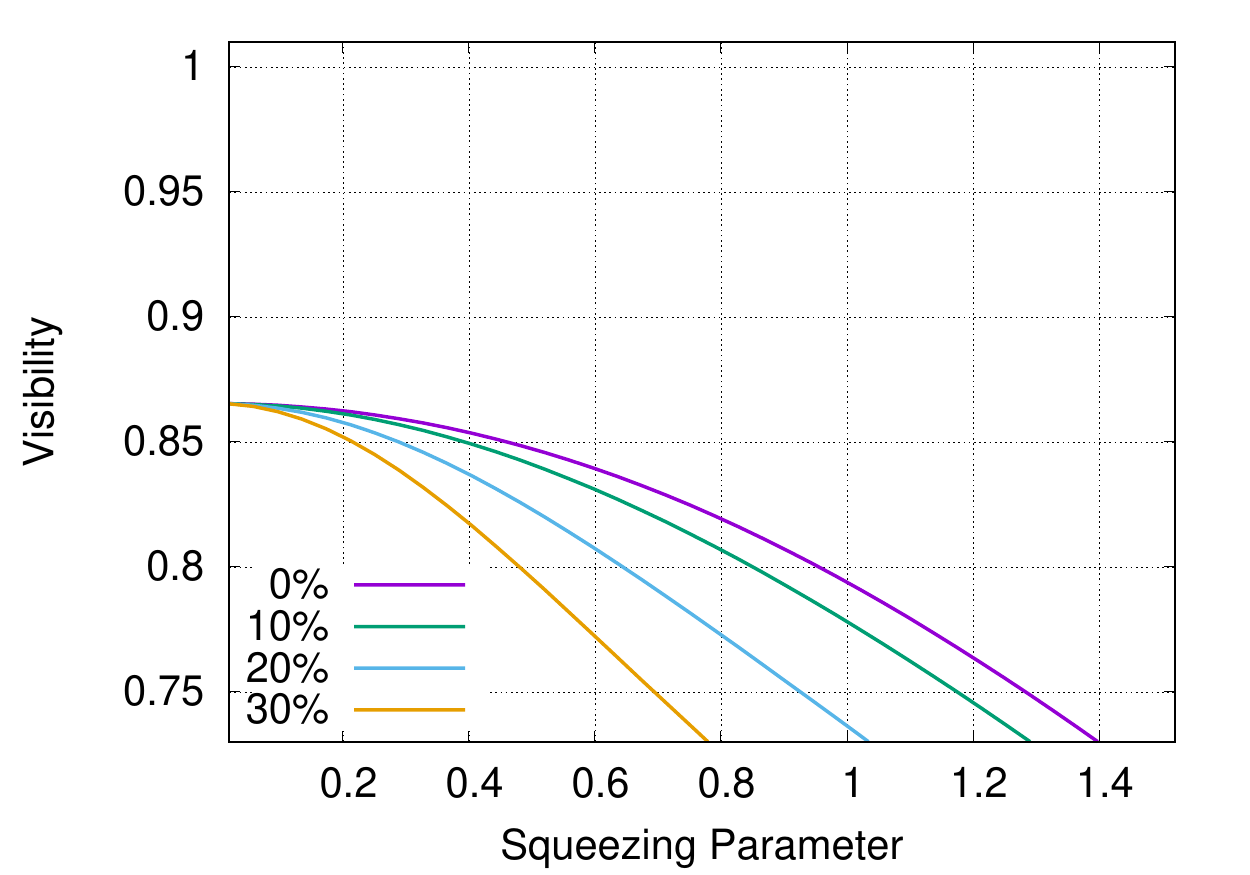}
\end{subfigure}
\caption{Interference visibility using number resolving detectors as a function of squeezing parameter (power), with each curve corresponding to different loss values as indicated. Part a) corresponds to two pure sources with separable JSAs, and part b) corresponds to sources with non-separable JSAs. Parameters as in Fig.~{\ref{fig:pure}} for part a) and Fig.~{\ref{fig:sinchomvis}} for part b).}
\label{fig:visloss}
\end{figure}

While the above analysis in the Fock basis does allow the state produced by a source to be scrutinised in this way, it is not a straightforward matter to describe its evolution through subsequent optical elements. As a pertinent example of this, and one of the main advantages of the formalism presented in this work, we now investigate the effect of frequency selective loss, i.e. filtering. The results so far have demonstrated that non-separability of the JSA is a significant factor affecting source figures of merit, and it is natural to ask to what extent separability of a JSA can be imposed by filtering.

In order to gain insight let us first consider the case in which a fixed amount of frequency independent loss is present on all modes, and investigate the measured interference visibility when using number resolving detectors. The results are shown in Fig.~(\ref{fig:visloss}), with parts a) and b)  corresponding to, respectively and as above, pure sources and non-separable sources. For each the different curves correspond to different loss levels as indicated. We see that with loss there is a greater decrease in visibility with increasing power as compared to the cases without loss. In fact, we see from part a) that even in the case for which the sources are pure and number resolving detectors are used, when loss is included there is a decrease in interference visibility with increasing power, which is not the case without loss. In all cases, the detrimental effects of loss can be understood as compromising the ability for a number resolving detector to herald a truly single photon state in the idler modes. 

So far the rate or efficiency of the sources investigated has been characterised by the heralding rate, which is the probability that there is a detection event in both of the heralding (signal) modes. In the absence of loss this quantity is precisely the probability that photon(s) are heralded in both of the idler modes since the photon numbers in the signal and idler modes are perfectly correlated. With the inclusion of loss, however, this is not the case, and it is instructive to also consider the heralding efficiency, defined as the conditional probability of detection events in both of the heralded (idler) modes $\mathit{(2,3)}$ when no beam-splitter is present, given both the herald (signal) modes $\mathit{(1,4)}$ registered a detection event. In our case this quantity can be calculated as the ratio of fourfold coincidence events measured with no interference to the heralding rate, and this is therefore written
\begin{equation}
    \eta^\mathrm{Herald}_{D} = \frac{P^\mathrm{SPS}_D}{P^\mathrm{Herald}_D},
    \label{eq:hefficiency}
\end{equation}
where $P^\mathrm{SPS}_\mathrm{D} = P^4_\mathrm{D} + P^\mathrm{bunch}_\mathrm{D}$ is the sum of the bunching and antibunching terms and is independent of the beam-splitter angle. Writing the heralding efficiency in this way allows it to be consistently defined even in the case where the beam-splitter angle is fixed.

\begin{figure}[t!]
\centering
\begin{subfigure}{0.48\textwidth}
\caption{}
\vspace{-0.4cm}
\includegraphics[width=1\textwidth]{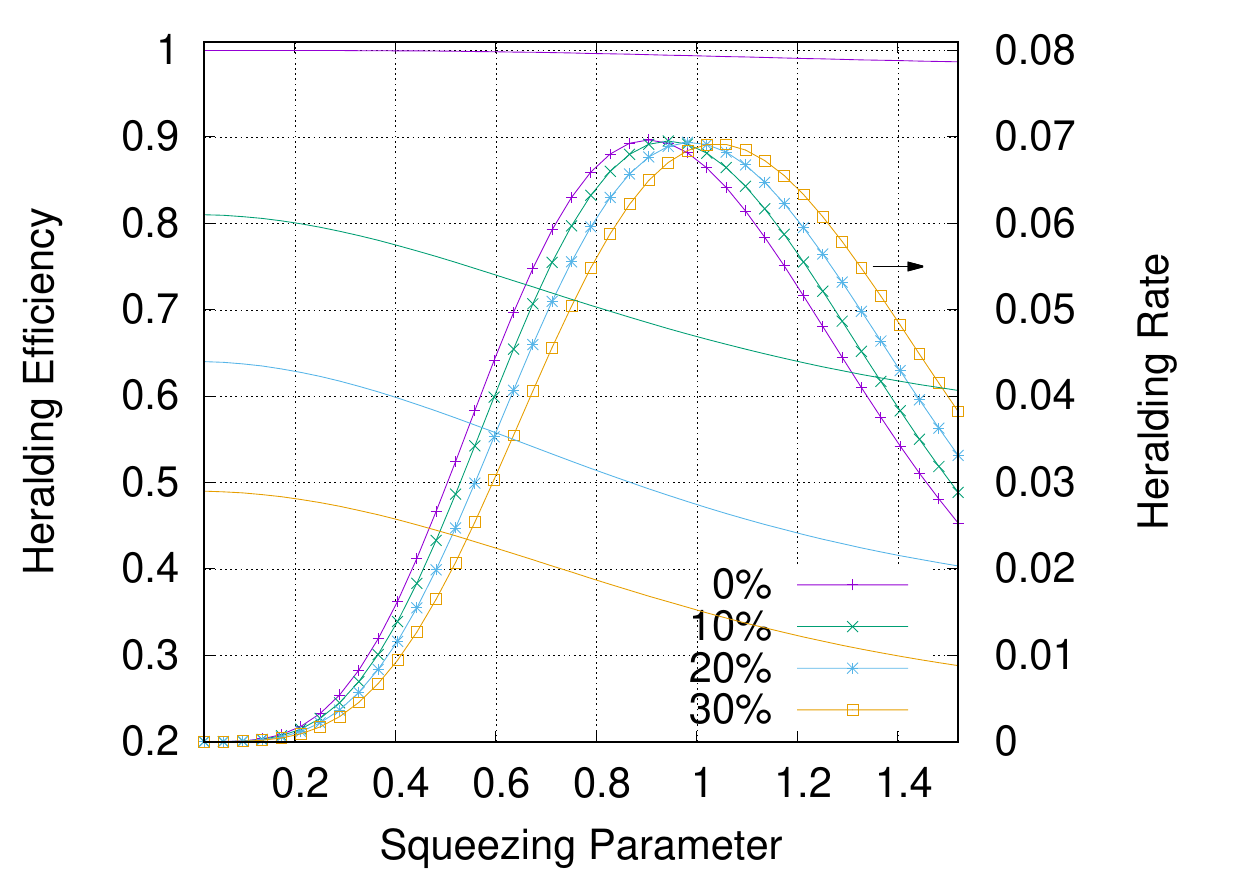}
\end{subfigure}
\hfill
\begin{subfigure}{0.48\textwidth}
\centering
\caption{}
\vspace{-0.4cm}
\includegraphics[width=1\textwidth]{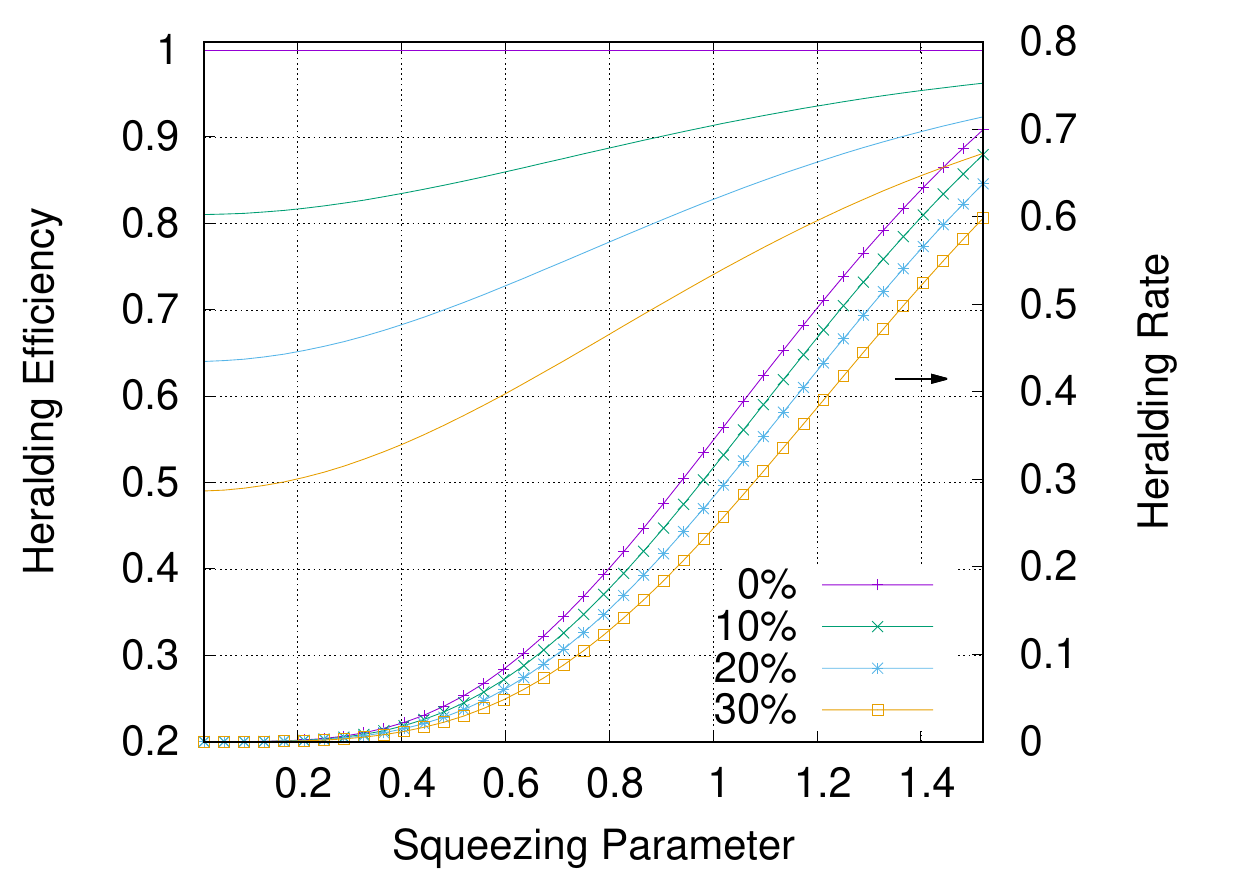}
\end{subfigure}
\caption{Heralding rates and efficiencies when using part a), number resolving detectors and part b), threshold detectors. The curves indicate the heralding efficiencies with different values of loss as indicated, and the plot points indicate the corresponding heralding rates.}
\label{fig:heralding}
\end{figure}

In Fig.~{\ref{fig:heralding}} we plot the heralding rate and heralding efficiency when including loss as a function of the squeezing parameter for both a) number resolving detection and b) threshold detectors, and in both cases consider non-separable sources with the parameters in Fig.~{\ref{fig:sinchomvis}}. We see that the peak in number resolved heralding rate is shifted to higher squeezing values with increasing loss, suggesting that decreases in photon generation rates causes by losses can in principle be easily overcome by increasing the pump power. However, while Fig.~{\ref{fig:visloss}} demonstrates that this compensation will necessarily decrease the interference probability, Fig.~{\ref{fig:heralding}} further shows that increasing the power will decrease the heralding efficiency, meaning that photons are less likely to be present in the idler modes when heralding events are registered. Loss therefore detrimentally affects the effective purity, heralding rate, and heralding efficiency of a source, with none of these figures of merit independently compensated by changing the pump power. With threshold detectors we similarly see that loss-induced decreases in the heralding rate can be compensated by increasing the squeezing parameter. Although the heralding efficiency appears to increase with increasing power for threshold detectors, this reflects only the fact that multiphoton terms increasingly tend to saturate these detectors.

\begin{figure}[t!]
    \centering
    \hfill
\begin{subfigure}{0.328\textwidth}
\centering
\vspace{-0.3cm}
\includegraphics[width=1.1\textwidth]{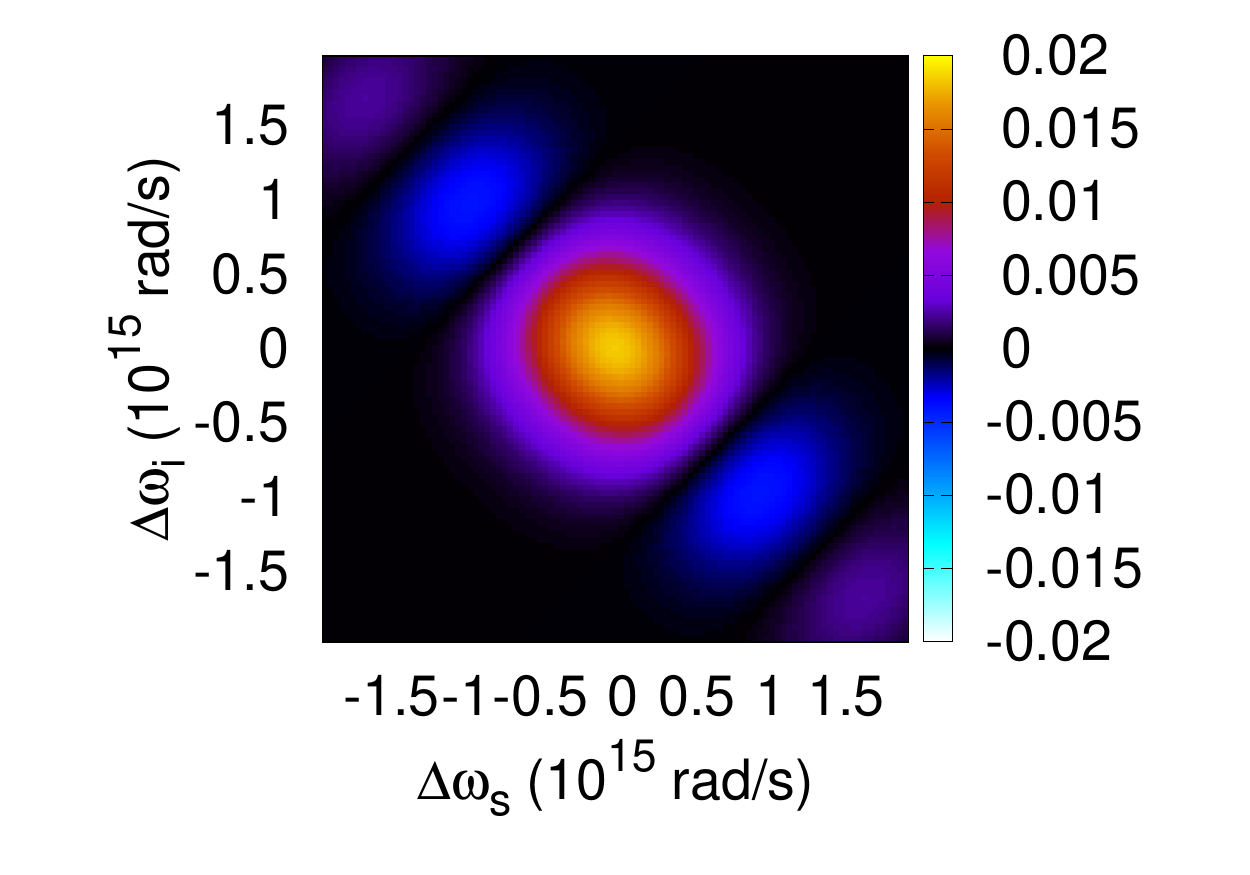}
\end{subfigure}
\begin{subfigure}{0.328\textwidth}
\centering
\vspace{-0.3cm}
\includegraphics[width=1.1\textwidth]{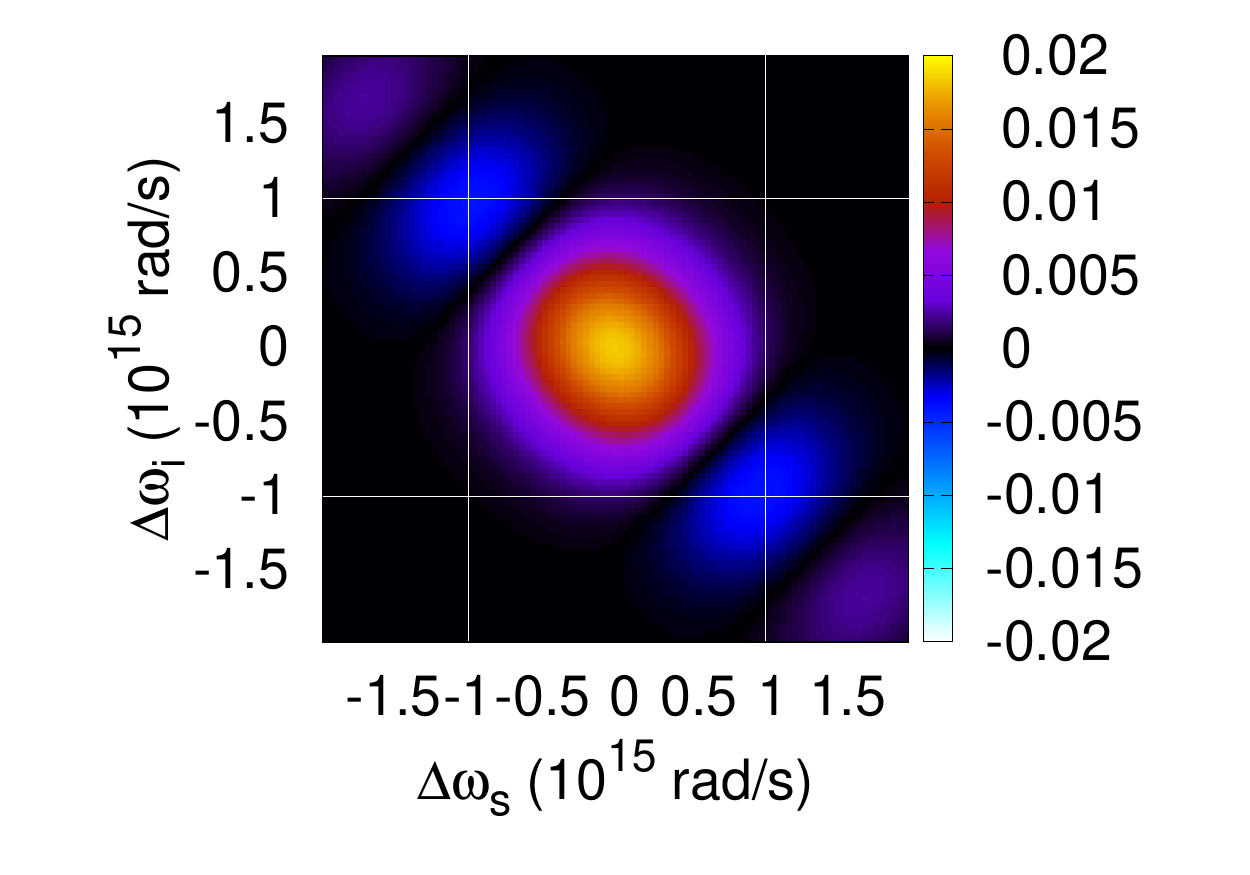}
\end{subfigure}
\hfill
\begin{subfigure}{0.328\textwidth}
\centering
\vspace{-0.3cm}
\includegraphics[width=1.1\textwidth]{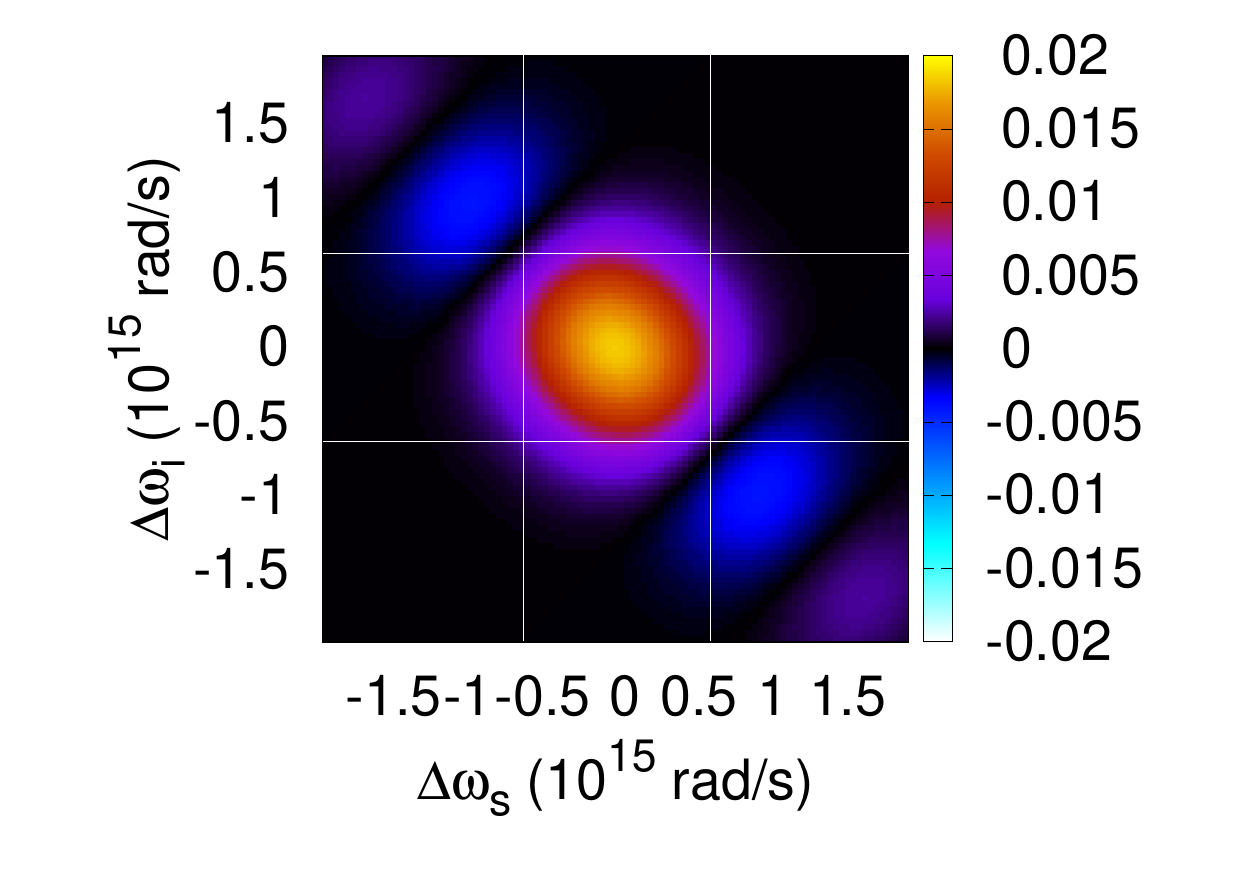}
    \end{subfigure}
    \includegraphics[width=\textwidth]{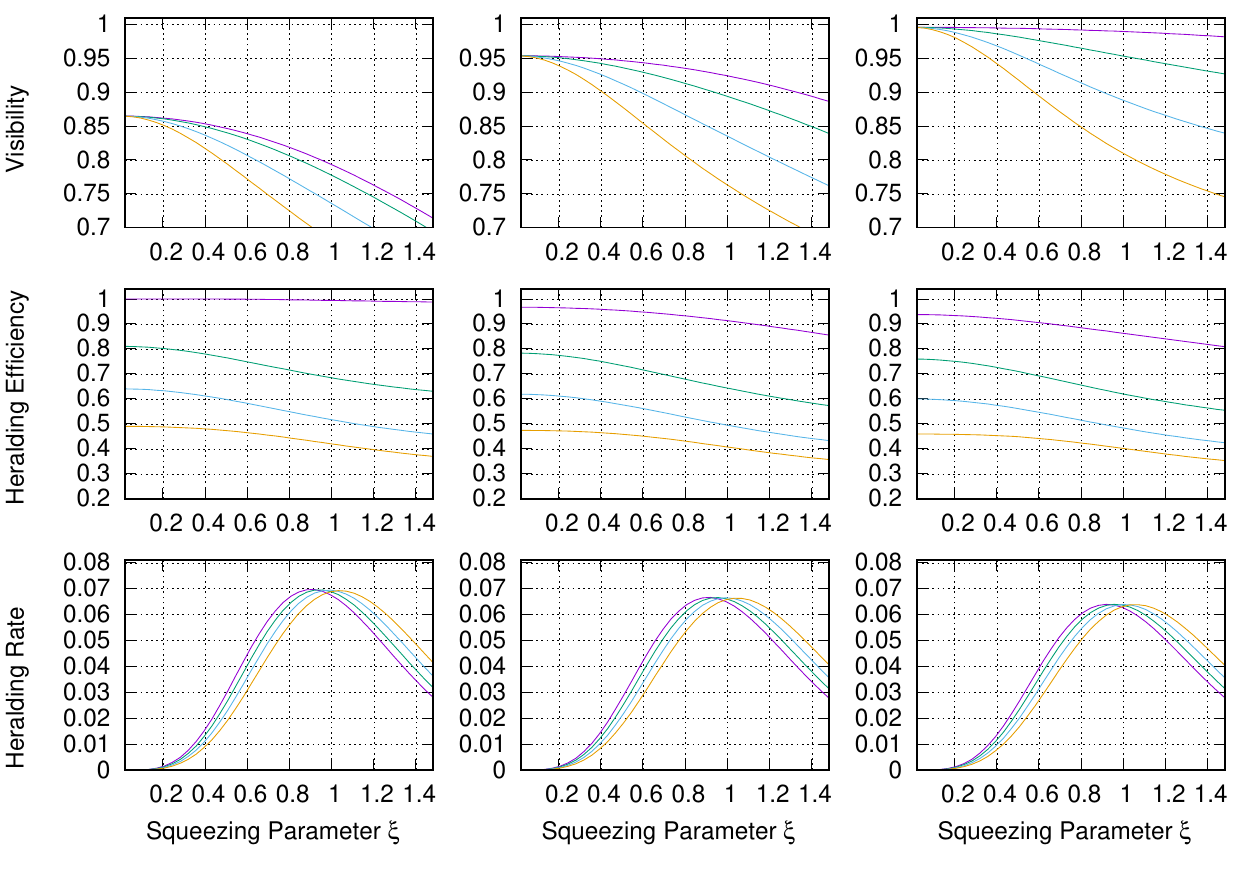}
    \caption{Source figures of merit when using number resolving detectors and no filtering (left column), a band-pass filter with width $2 {\times} 10^{15}$~rad/s (middle column), and similarly with width $1.2 {\times} 10^{15}$~rad/s (right column). Unfiltered source has a non-separable JSA as in Fig.~{\ref{fig:sinchomvis}}, and the different curves correspond to different loss levels as in Fig.~{\ref{fig:heralding}}.
    We see that tight filtering (right column) achieves near unit visibility in the low power limit, however this does not persist with increasing power, and becomes very sensitive to loss.}
    \label{fig:sincfilter}
\end{figure}
Let us now consider the effect of spectral filtering. We investigate the case in which two sources each described by a non-separable JSA are subject to spectral filtering on both spatial modes (signal and idlers). Each mode is filtered with the same ideal bandpass filter through which photons are either fully transmitted or fully removed depending on their frequency. The bandpass filters are centred at the centre of the JSAs ($\nu_0=\bar{\nu}=193.1~\mathrm{THz}$) and the sources have the same parameters as in Fig.~{\ref{fig:sinchomvis}}. The results are shown in Fig.~{\ref{fig:sincfilter}}, where the left column corresponds to no filtering, the middle column to a filter width of $\Delta\nu_f=2{\times}10^{15}$~rad/s, and the right column to $\Delta\nu_f=1.2{\times}10^{15}$~rad/s. The widths of the filers are indicated by the white lines in the plots of the real part of the JSA in the top row, and in all cases the different curves in the plots below correspond to the different levels of loss. As one might expect, we see that tight filtering of the sources means that at low powers an interference visibility of up to 99.58\% can be achieved for the initially non-separable imperfect source. The trade off is that the filtering process has reduced the maximum heralding rate from 0.07 to 0.063, and the heralding efficiency decreases to from 100\% to 88\% at the point of maximal heralding rate.

More importantly, however, is that it appears spectral filtering of this sort does not allow unit visibility for all squeezing parameter values, even with number resolving detectors; a tightly filtered source does not behave like a separable source, even if perfect interference visibilities are found in the low power regime. For a truly pure source, as explored in Fig.~{\ref{fig:pure}}, the use of number resolving detectors allows the heralding rate to be increased by increasing the pump power at no cost to the interference visibility. In the present case, however, this is no longer true, even when there is no (frequency independent) loss. Once again, the problem here is that filtering has introduced photon number noise, and number resolving detection is no longer able to herald a truly single photon state.

\begin{figure}[t]
\centering
\begin{subfigure}{0.328\textwidth}
\includegraphics[width=\linewidth]{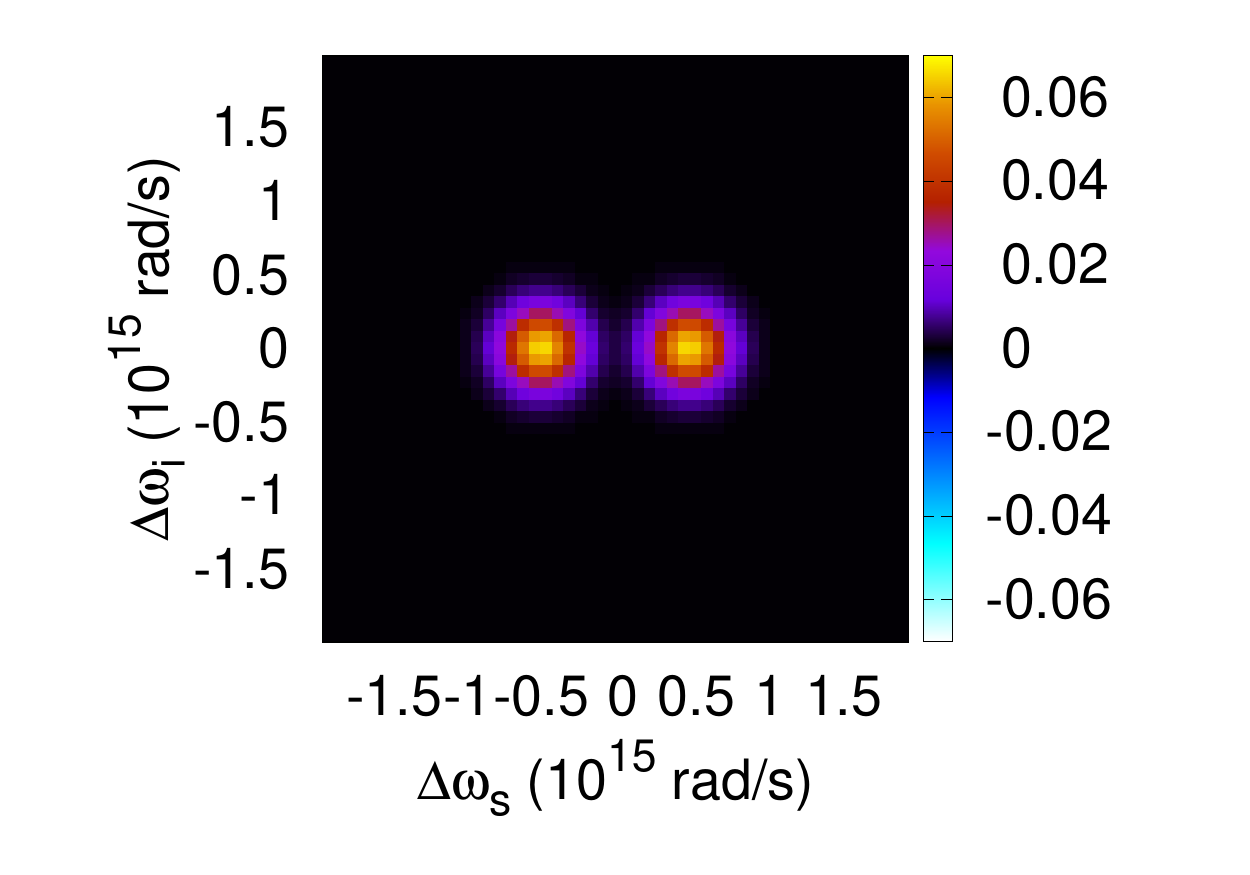}
\end{subfigure}
\hfill
\begin{subfigure}{0.328\textwidth}
    \includegraphics[width=\linewidth]{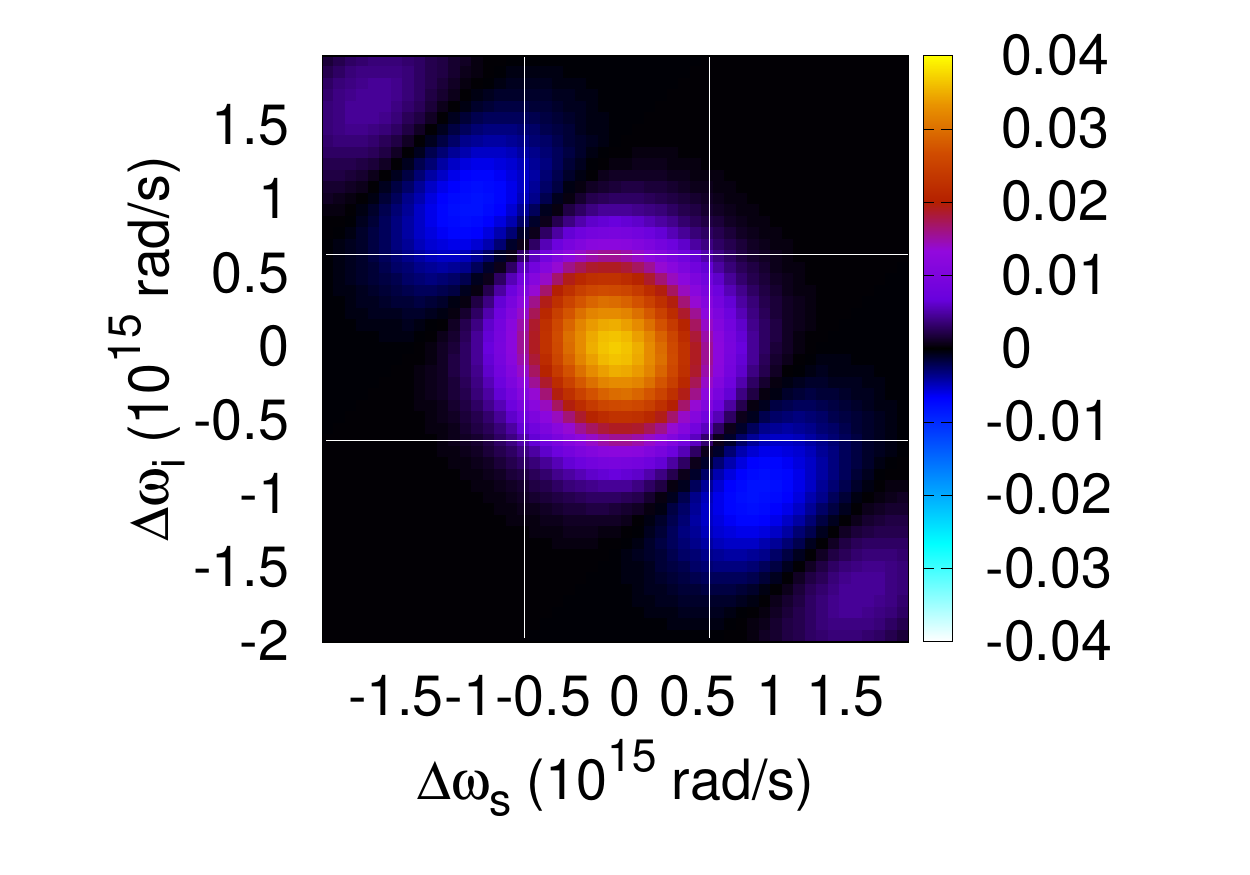}
\end{subfigure}
\hfill
\begin{subfigure}{0.328\textwidth}
\includegraphics[width=\linewidth]{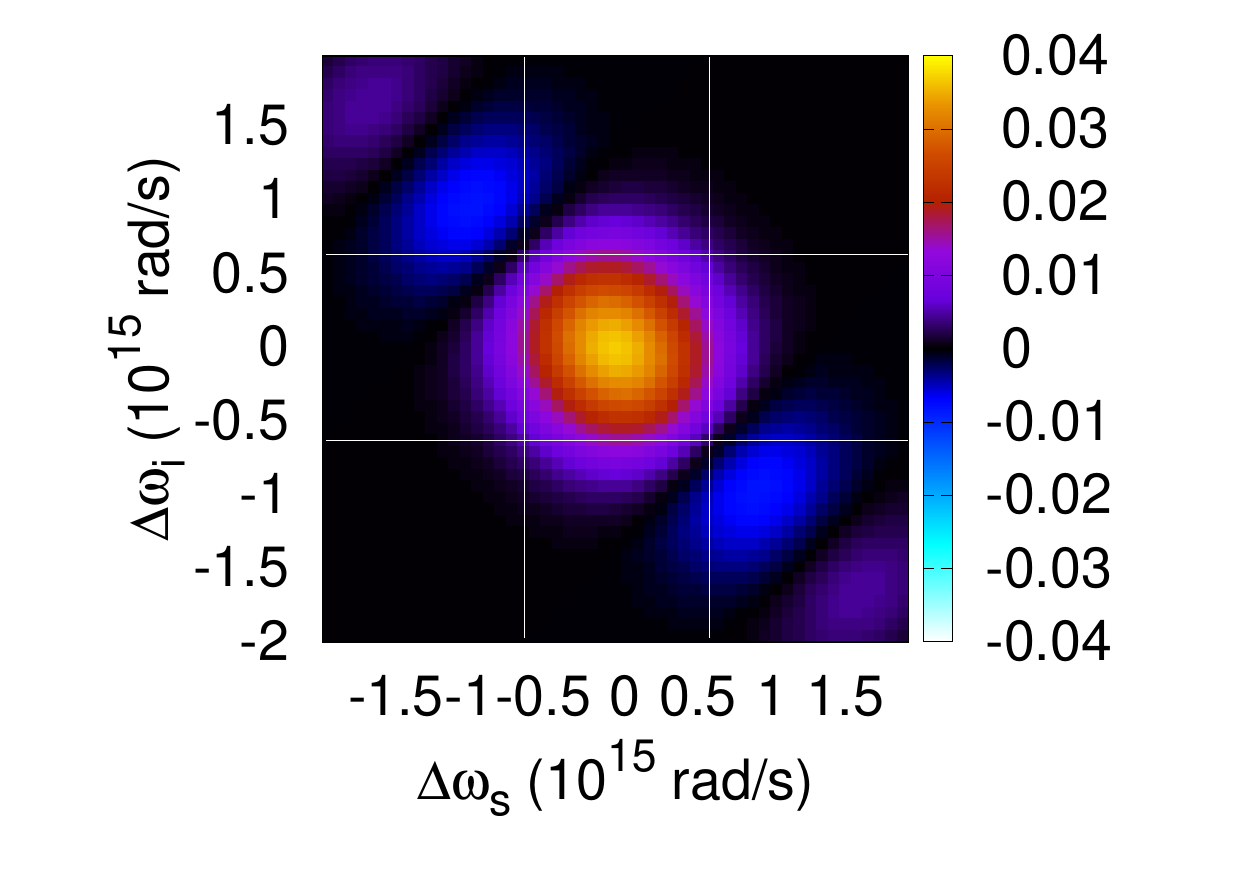}
\end{subfigure}

\begin{subfigure}{0.328\textwidth}
\includegraphics[width=\linewidth]{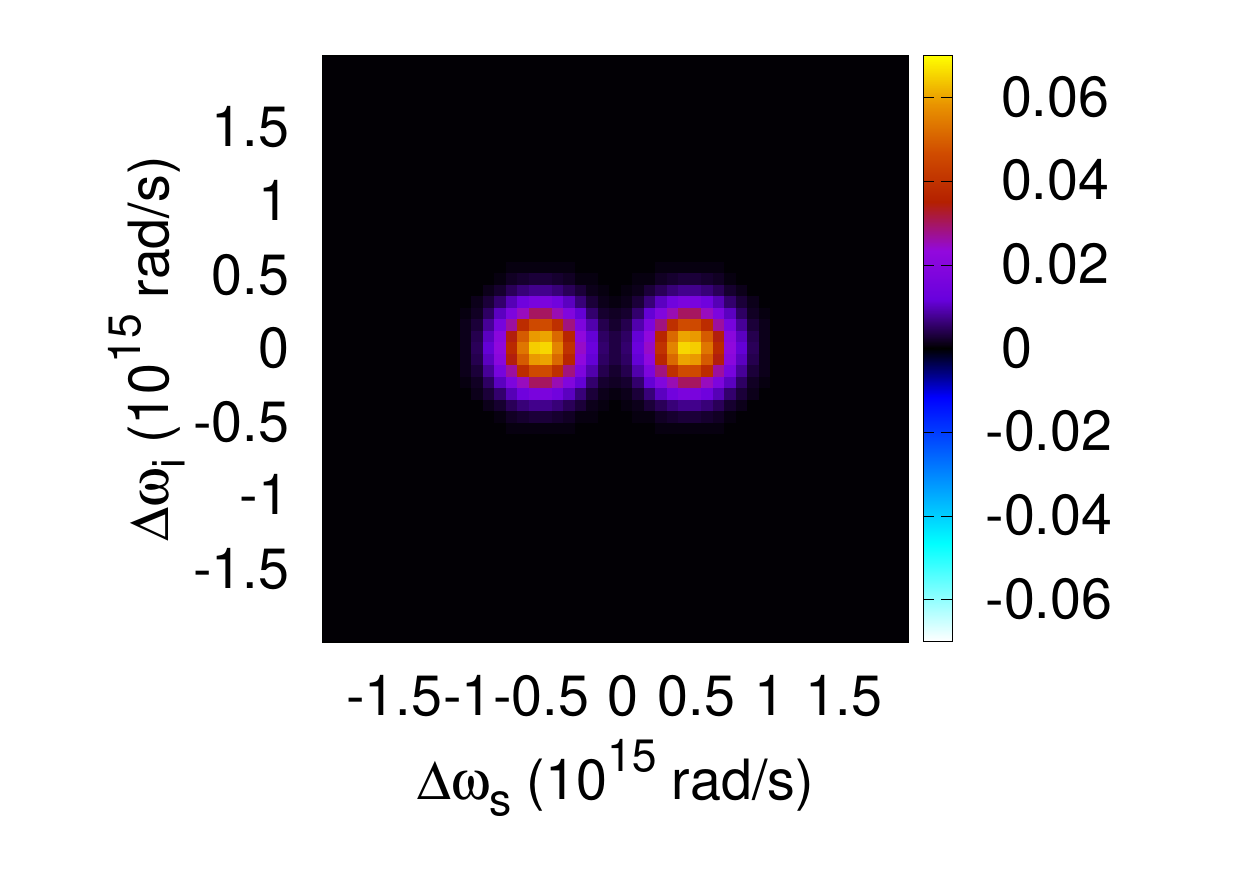}
\end{subfigure}
\hfill
\begin{subfigure}{0.328\textwidth}
    \includegraphics[width=\linewidth]{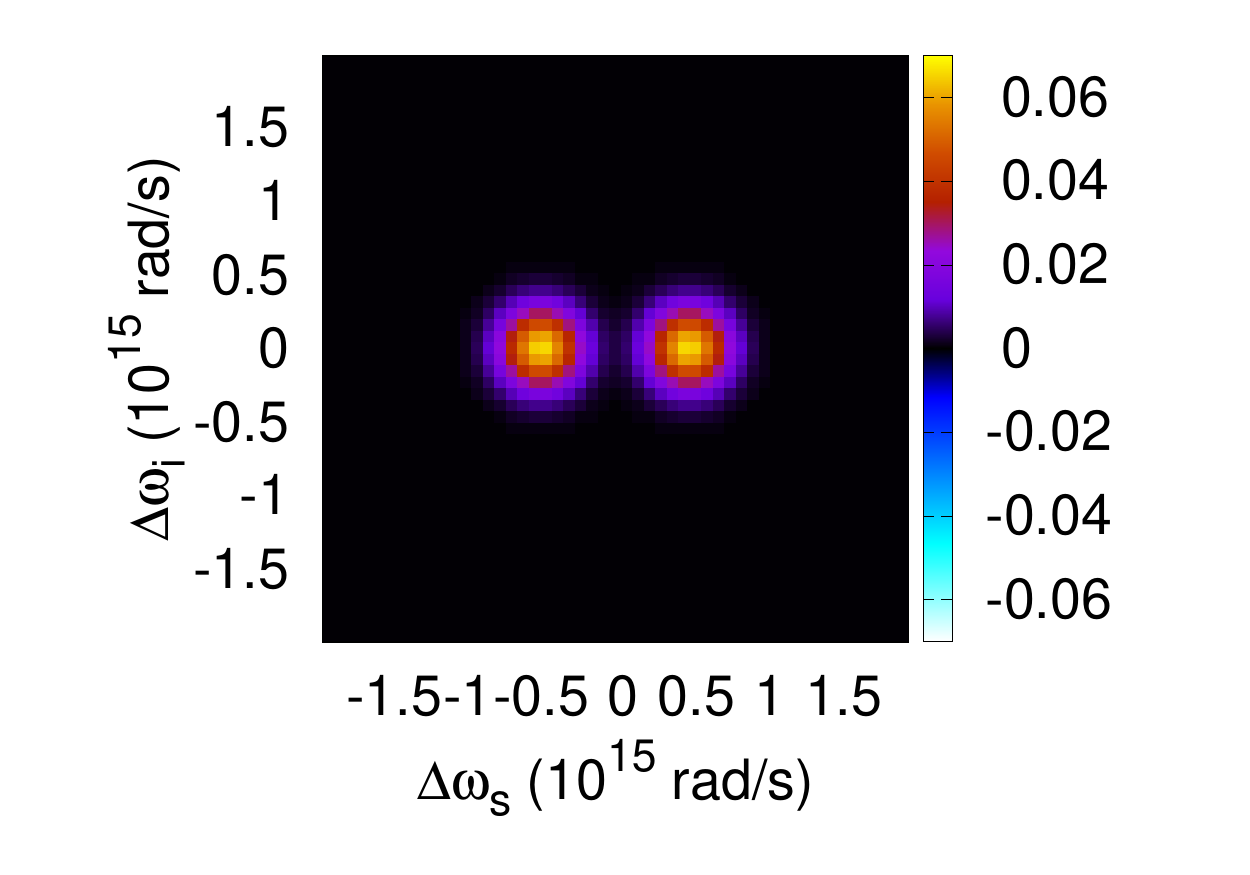}
\end{subfigure}
\hfill
\begin{subfigure}{0.328\textwidth}
    \centering
    \includegraphics[width=1\linewidth]{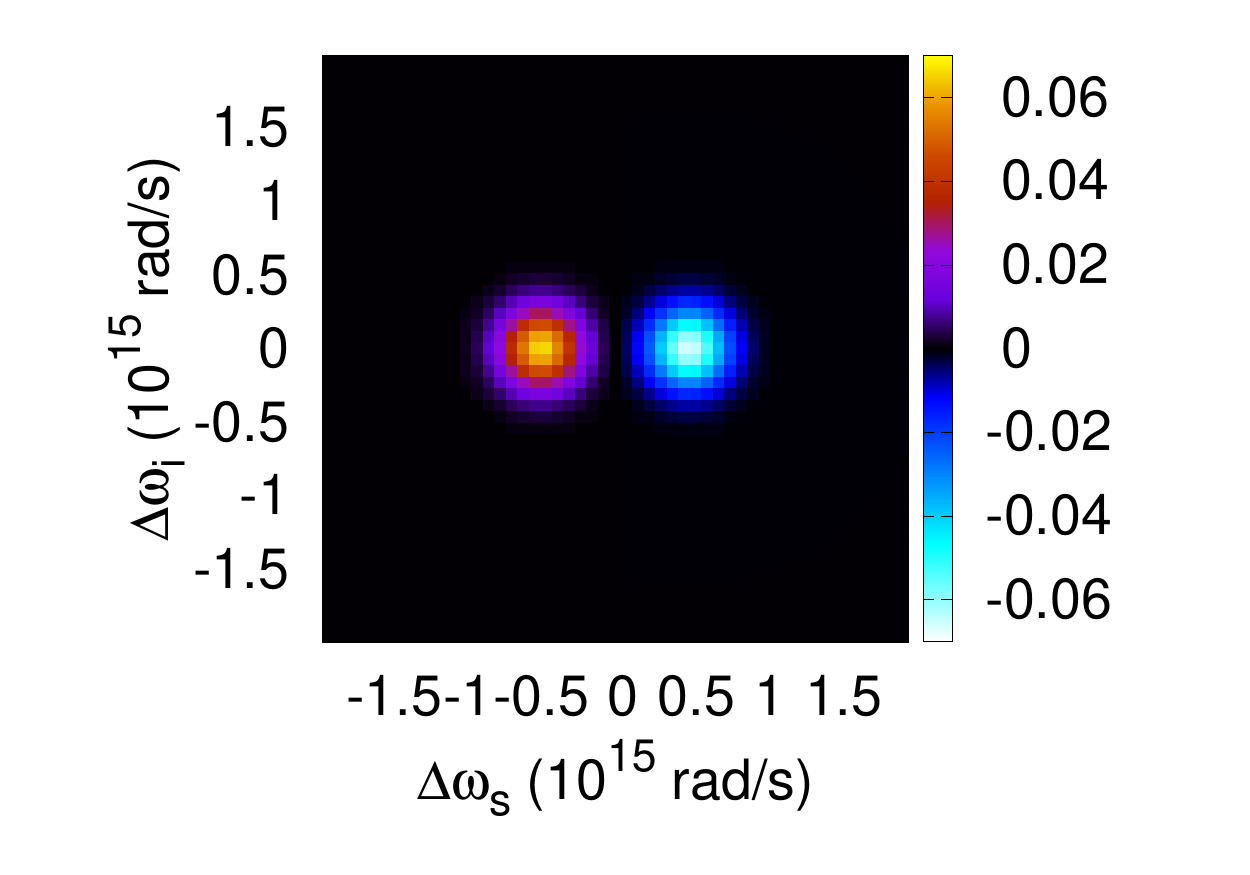}
\end{subfigure}

\begin{subfigure}{0.31\textwidth}
\includegraphics[width=\linewidth]{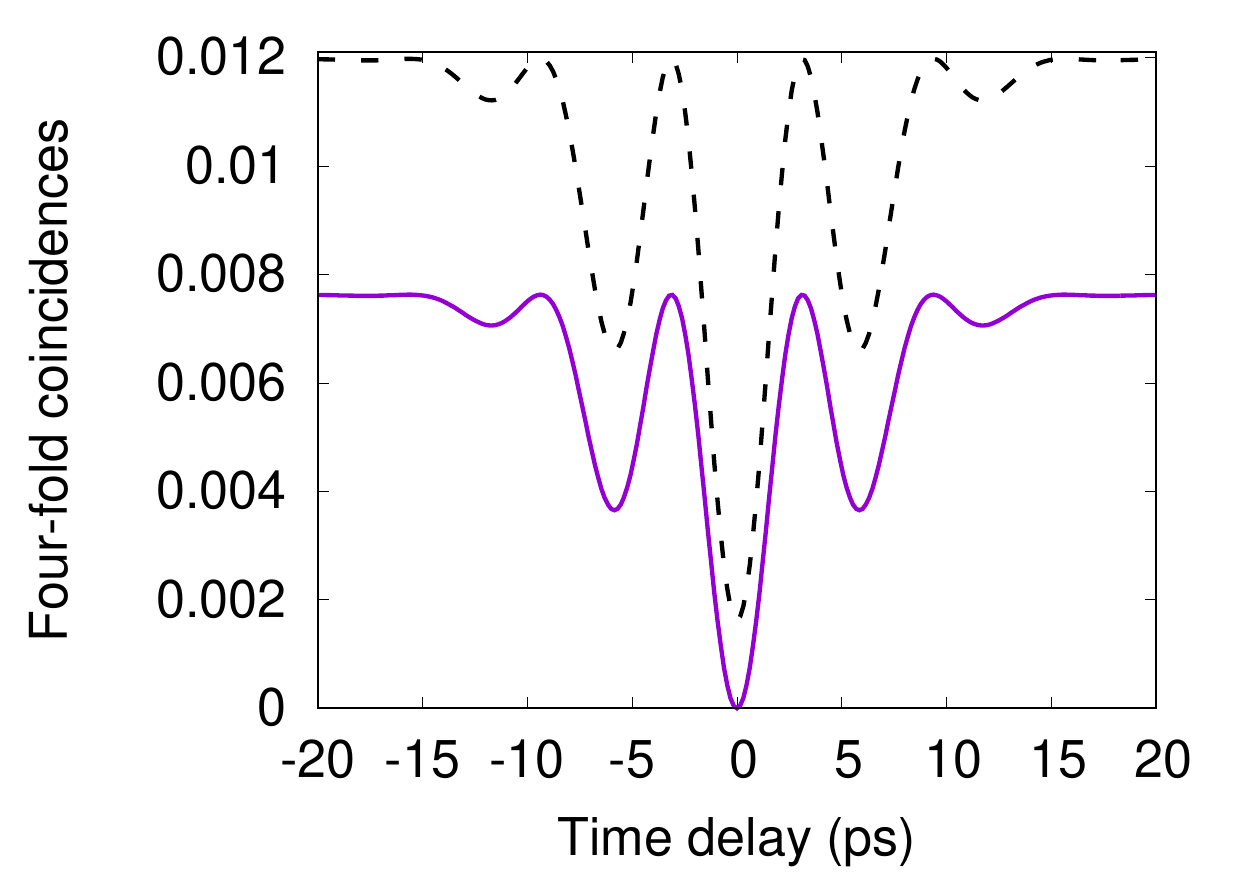}
\end{subfigure}
\hfill
\begin{subfigure}{0.31\textwidth}
    \includegraphics[width=\linewidth]{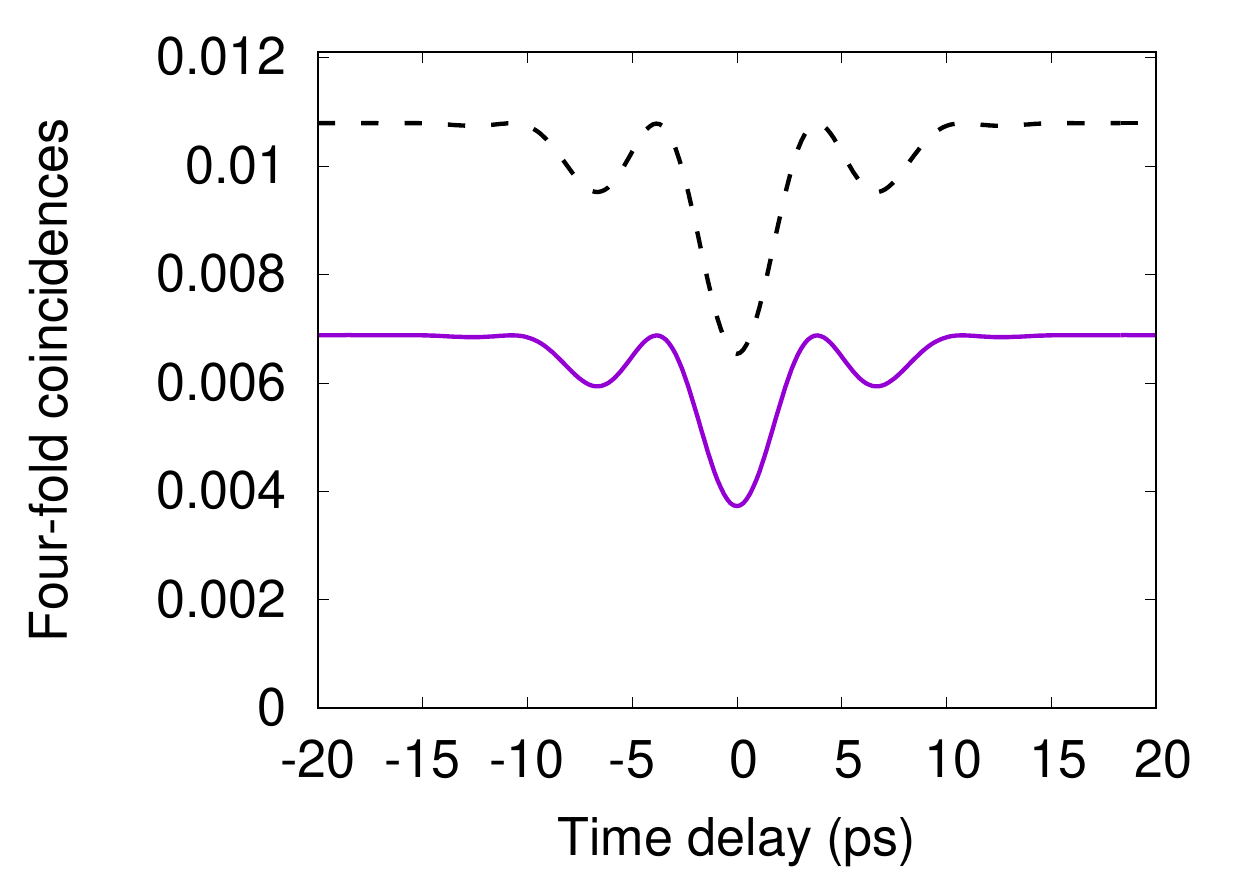}
\end{subfigure}
\hfill
\begin{subfigure}{0.31\textwidth}
    \centering
    \includegraphics[width=1\linewidth]{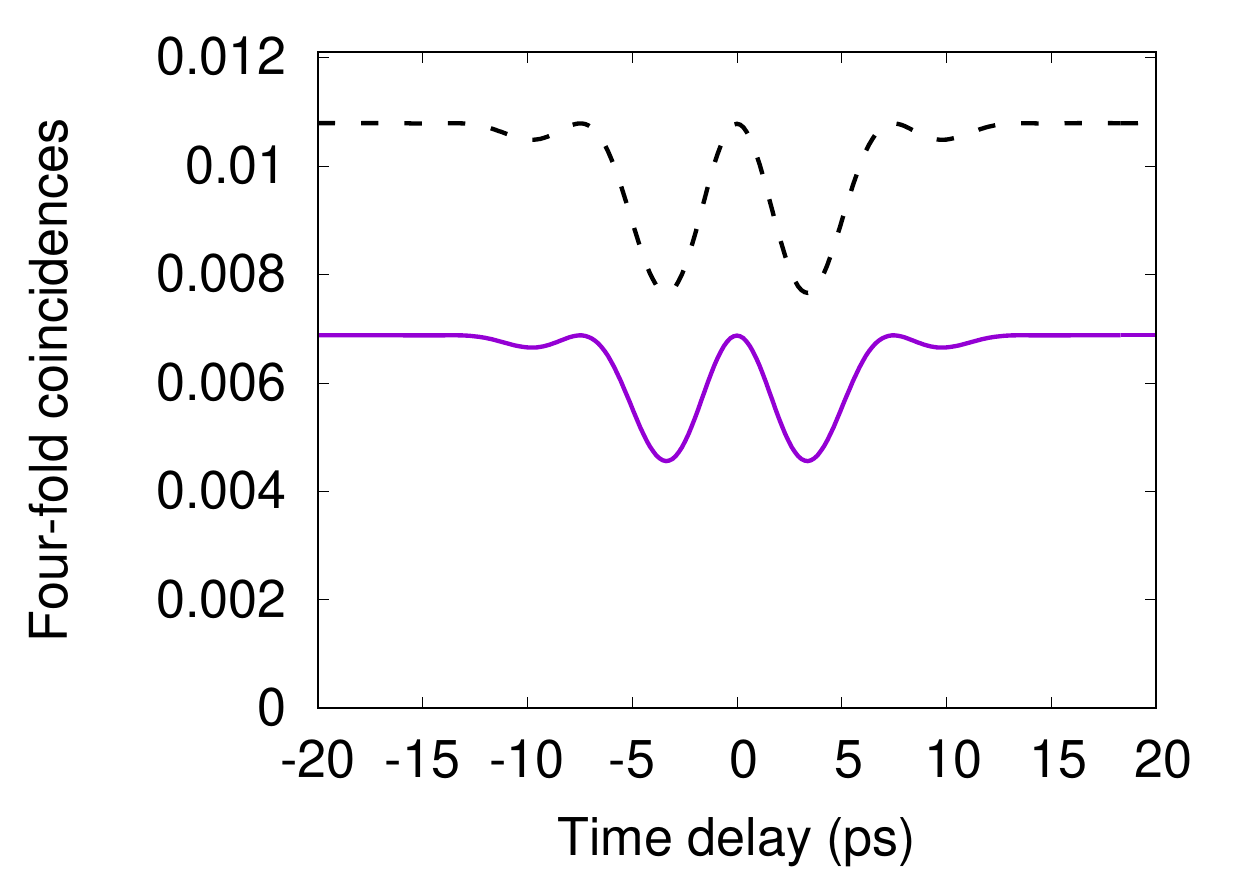}
\end{subfigure}

\caption{Variable time delay HOM measurements for more exotic sources. The left columns shows the real part of two identical and separable sources, each consisting of two Gaussian lobes with width $0.03$~THz. The middle column corresponds to a filtered non-separable waveguide source as in Eq.~({\ref{spiralJSA}}), and a separable double lobed source. The right column shows a filtered non-separable source and a double lobed source with each lobe of opposite sign. All sources have squeezing parameter of $\xi=0.4$. The bottom row shows the associated four-fold coincidences as measured using threshold detectors, $P^4_\mathrm{Thres}$ (black, dashed), and number resolving detectors, $P^4_\mathrm{PNR}$ (purple).}
\label{StructuredSources}
\end{figure}

\subsection{Structured and non-identical sources}

As a final demonstration of the utility of our presented formalism, we will now consider heralded HOM interference between structured and non-identical sources. The pairs of sources that we will consider are shown in the first two rows of Fig.~{\ref{StructuredSources}}. The first column shows a pair of identical sources, with the JSA of each consisting of the sum of two Gaussian functions of the form in Eq.~({\ref{eq:gaussjsa}}) with widths $\zeta=0.03~\mathrm{THz}$. The four-fold coincidences are shown in the bottom plot, calculated for number resolving detectors (solid, purple curve) and threshold detectors (dash, black). The heralded photons here have double peaked temporal (and spectral) envelops, resulting in a HOM dip with multiple features reflecting the Fourier transform of the product of the two heralded photon spectra. In the middle column we keep one source the same, and replace the other with a filtered waveguide source with a non-separable JSA of the form in Eq.~({\ref{spiralJSA}}). The corresponding HOM dip shown in the plot at the bottom again has interesting temporal features, and here the dip minima are higher than in the previous case, reflecting both the spectral impurity of the first source and the poor overlap of the two heralded photons. Finally, in the column on the right we consider again the filtered non-separable waveguide source, now interfering with a double peaked Gaussian with opposite phase, i.e. a JSA that is the difference of two terms of the form in Eq.~({\ref{eq:gaussjsa}}). Here we see almost no interference at $0$ time delay, as the two heralded photons now have approximately orthogonal spectra. However, for time delays of $\Delta\tau\approx\pm 4~\mathrm{ps}$ we see the interference is improved, as for these parameters one of the peaks of the second source is centred on that of the first. We stress that in all cases here, while it may be possible to write down the conditional state of the heralded photons in the Fock basis and describe the qualitative features of these HOM dips, calculating the actual four-fold coincidence count rates and deriving a corresponding interference visibility beyond the weak excitation limit is considerably more involved. Within our continuous variable formalism, on the other hand, the complete state of the system to all photon number orders and with all spectral degrees of freedom is explicitly propagated through all optical elements, allowing for photon detection rates to be calculated exactly.

\section{Conclusion}

In this work we have introduced a Gaussian quantum optics formalism that explicitly includes a tensor product structure of the spatial and spectral modes. This formalism allows us to model realistic and relevant highly multimoded fields generated by collections of parametric sources, together with non-Gaussian threshold and number resolving detection. This has allowed us to extend the known expressions for non-frequency resolving threshold and number resolved detection probabilities to include spectral degrees of freedom. We applied this formalism to the study of heralded Hong--Ou--Mandel interference, in which we elucidated the inter-dependencies of the source heralding rates, heralding efficiencies and interference visibilities. In particular, we demonstrated that any non-separability of the source joint-spectral amplitude results in decreases of the interference visibility beyond the weak excitation limit, and further decreases in the interference visibility and heralding efficiency when any loss is present. Furthermore, we found that while spectral filtering can improve a source's interference visibility, the filtering process necessarily introduces photon number noise which detrimentally affects the interference visibility and heralding efficiency at higher powers.

The HOM setup considered here constitutes a fundamental primitive for many more complex systems, and as such understanding these effects will become ever more important as demonstrations (and applications) are increased in scale. Consider for instance, current state-of-the-art 4-photon experiments achieving detection rates of $10^{-2}~\mathrm{s}^{-1}$~\cite{Adcock2019}, $1~\mathrm{s}^{-1}$~\cite{Sun2019,Vigliar2020}, $1-2~\mathrm{s}^{-1}$~\cite{McCutcheon2016}, $21~\mathrm{s}^{-1}$~\cite{Li2019a}, and $46~\mathrm{s}^{-1}$~\cite{Proietti2019}, with target state fidelities not exceeding $0.94$. 
Whilst state infidelity is routinely attributed to multi-photon events, with heuristic or idealised error models of such often presented~\cite{Graffitti2017}, an exact analysis including multiphoton events alongside spectral effects and detection mechanisms has yet to be included. 
As systems are required to provide higher count rates, source figures of merit including purity and multi-photon contamination are traded in favour of achieving acceptable count rates. Recently a 10-photon experiment was demonstrated at a rate 4 per hour~\cite{Wang2016} at the expense of a state fidelity of $0.573$, and a subsequent 12-photon experiment achieved fidelity of $0.576$ at a rate 1 per hour~\cite{Zhong2018}. Of course, the more efficient generation of higher photon-number states for computation applications requires multiplexing~\cite{Kaneda2019}, and one would necessarily have to chose an operation regime in which generation rates are traded off against multi-photon contamination, and a realistic noise model capturing multi-photon and spectral impurity is then required in order to demonstrate that acceptable error thresholds are indeed overcome. In particular, it is clear that leakage errors, associated to the higher dimensional state space of multi-photon states (outside the desired photon number subspaces) will present a key challenge for both the experimental systems, and the theory of their operation~\cite{Wood2017}.  

In this work we have primarily focused on the generation of heralded single photon states, which could in turn to be used to generate larger states with fixed photon number. However, our formalism also naturally lends itself to applications where states without fixed photon number are of interest, most notably those of Gaussian boson sampling, which are based on the observation that calculating photon detection probabilities from Gaussian states with many modes is computationally hard~\cite{quesada2018gaussian,Paesani2019,zhong2020quantum}. Using the present formalism it would be interesting to investigate the extent to which photon spectral impurity or mutual distinguishability can affect the computational complexity of multi photon detection probabilities. It is in turn worth mentioning that the number resolving detection probabilities derived here make use only of derivatives of vacuum projections and combinatorics, and it would be interesting to explore the extent to which these new expressions could lead to insights into the development of classical algorithms for simulating photonic measurements of this sort.

\ack
O.F.T. was supported by the Bristol Quantum Engineering Centre for Doctoral Training, EPSRC grant EP/L015730/1. We would like to thank J. R. Scott for useful discussions, and S. E. Armstrong and J. F. F. Bulmer for their helpful comments.

\newcommand{\newblock}{}
\section*{References}
\bibliographystyle{unsrt}
\bibliography{references}

\begin{appendices}

\section{Characteristic functions and symplectic transformations}

We define the $s$-ordered characteristic function as 
\begin{equation}
\chi_{\hat{\rho}}^{(s)}(\vec{\Lambda})=\mathrm{Tr}[\hat{\rho} D(\vec{\Lambda})]\exp[\sfrac{1}{4} s |\vec{\Lambda}|^2],
\end{equation}
where $\hat{\rho}$ is the corresponding quantum state, $D(\vec{\Lambda})=\smash{\exp \big[\hat{\vec{A}}^{\dagger} K \vec{\Lambda}\big]}$ with $\hat{\vec{A}}$ given in Eq.~({\ref{Aspectralandspatial}}), and $\vec{\Lambda}$ is a $2N$-dimensional vector of complex variables. Note that this vector takes the general form $\vec{\Lambda}=(\vec{\lambda},\vec{\lambda}^*)^\top$ with $\vec{\lambda}=(\lambda_{1\w_1},\dots,\lambda_{N})^\top$
\footnote{With these definitions which can include, for example, both $\vec{\lambda}$ and $\vec{\lambda}^*$ in $\vec{\Lambda}$, we need to be careful when considering absolute values. In particular, we note that $|\vec{\Lambda}|^2=\vec{\Lambda}^{\dagger}\vec{\Lambda}=2|\vec{\lambda}|^2=2\sum_{i\w} |\lambda_{i\w}|^2=\sum_{i\w} |\vec{\lambda}_{i\w}|^2$.}. To see how a quantum state and corresponding characteristic function evolve under a unitary transformation, we suppose that the state transforms as $\hat{\rho}\to\hat{\rho}'=\hat{U}\hat{\rho}\hat{U}^{\dagger}$. The corresponding characteristic function is 
\begin{equation}
    \chi_{\hat{\rho}'}^{(s)}(\vec{\Lambda})=\mathrm{Tr}[\hat{U}\hat{\rho}\hat{U}^{\dagger} D(\vec{\Lambda})]\exp[\sfrac{1}{4} s |\vec{\Lambda}|^2].
\end{equation}
For a linear symplectic transformation as described in Eq.~(\ref{MfromH}) we find 
\begin{equation}
    \chi_{\hat{\rho}'}^{(s)}(\vec{\Lambda})=\chi_{\hat{\rho}}^{(s)}(M^{-1}\vec{\Lambda}).
\end{equation}
For the sake of completeness we note that generalising Eq.~({\ref{MfromH}}) to affine transformations, $\hat{U}^{\dagger} \hat{\vec{A}} \hat{U}=M\hat{\vec{A}}+\vec{m}$, we have have 
\begin{equation}
    \chi_{\hat{\rho}'}^{(s)}(\vec{\Lambda})=\exp[\vec{m}^{\dagger}K\vec{\Lambda}]\chi_{\hat{\rho}}^{(s)}(M^{-1}\vec{\Lambda}).
    \label{SymplecticChiTransformation}
\end{equation}
Non-zero $\vec{m}$ contributions correspond to the injection of coherent states.

Gaussian states are defined as those states with characteristic functions which are Gaussian. A general Gaussian state therefore takes the form
\begin{equation}
\chi_{\hat{\rho}}^{(s)}(\vec{\Lambda})=\exp[-\sfrac{1}{4}\vec{\Lambda}^{\dagger} K (\sigma-s\mathds{1}_{2 N}) K \vec{\Lambda}+\vec{d}^{\dagger} K \vec{\Lambda}],
\label{ComplexGaussianChi}
\end{equation}
where the $2N\times 2N$ matrix of scalar coefficients $\sigma$ is known as the covariance matrix, and the $2N$-dimensional vector $\vec{d}$ is the displacement vector. If a Gaussian state undergoes a symplectic transformation described by Eq.~({\ref{SymplecticChiTransformation}}), using the symplectic condition in Eq.~({\ref{SymplecticCondition}}) it follows that the resulting characteristic function remains Gaussian, but with modified covariance matrix and displacement vector given by 
\begin{equation}
\sigma\to\sigma'=M\sigma M^{\dagger},\qquad \vec{d}\to\vec{d}'=M\vec{d}+\vec{m}.
\end{equation}

\subsection{Calculation of projection onto the vacuum}

In terms of characteristic functions, the expectation value of a general operator $\hat{O}$ is written
\begin{equation}
\langle \hat{O} \rangle = \frac{1}{\pi^N}\int d^{2N}\vec{\Lambda} \chi_{\hat{\rho}}^{(s)}(\vec{\Lambda}) \chi_{\hat{O}}^{(-s)}(-\vec{\Lambda}),
\label{ExpectationOComplexN}
\end{equation}
where $d^{2N}\vec{\Lambda}=\prod_{i\w} d\mathrm{Re}[\lambda_{i\w}]d \mathrm{Im}[\lambda_{i\w}]$. In what follows it will also be convenient to define the multimode quasi-probability distributions as 
\begin{equation}
W_{\hat{\rho}}^{(s)}(\vec{\mathcal{A}})=\frac{1}{\pi^{2N}}\int d^{2N}\vec{\Lambda} \chi_{\hat{\rho}}^{(s)}(\vec{\Lambda}) \exp[\vec{\Lambda}^{\dagger} K \vec{\mathcal{A}}].
\label{WsDefinitionN}
\end{equation}
The $s=-1,0,1$ forms are known as, respectively, the Husimi-Q function, Wigner function, and Glauber--Sudarshan P representation. A Gaussian state of the form in Eq.~({\ref{ComplexGaussianChi}}) has the quasi-probability distributions 
\begin{equation}
W_{\hat{\rho}}^{(s)}(\vec{\mathcal{A}})=\frac{\left(2/\pi\right)^N}{\sqrt{\mathrm{det}[\sigma-s\mathds{1}_{2N}]}}
\exp[-(\vec{d}-\vec{\mathcal{A}})^{\dagger}(\sigma-s\mathds{1}_{2N})^{-1}(\vec{d}-\vec{\mathcal{A}})],
\label{GaussianWsN}
\end{equation}
valid for $s=-1$ or $s=0$. A closed-form expression for the Glauber--Sudarshan P representation ($s=1$ form) for a general Gaussian state does not exist. 

In order to calculate projections of a general Gaussian state onto the vacuum for a subset of the modes, we label the subset $S$ and its complement $\bar{S}$. We are interested in the probability associated with the projector onto the vacuum for all modes in $S$, which we label $\ketbra{\mathrm{vac}}{\mathrm{vac}}_S$. For a general operator $\smash{\hat{O}_S}$ acting only on modes in $S$ we have 
\begin{equation}
\langle \hat{O}_S \rangle=\mathrm{Tr}_{S+\bar{S}}[\hat{\rho} \hat{O}_S \hat{\mathds{1}}_{\bar{S}} ]=\mathrm{Tr}_{S}[\hat{\rho}_S \hat{O}_S ]=\frac{1}{\pi^{|S|}}\int d^{2|S|} \chi^{(s)}_{\hat{\rho}_S}(\vec{\Lambda}_S)\chi^{(-s)}_{\hat{O}_S}(-\vec{\Lambda}_S)
\label{ExpectationOComplexS}
\end{equation}
where $\hat{\mathds{1}}_{\bar{S}}$ is the identity in the space of modes $\bar{S}$, 
$\hat{\rho}_S=\mathrm{Tr}_{\bar{S}}[\hat{\rho}]$ is the reduced density matrix on modes $S$, and 
$\vec{\Lambda}_S$ contains variables pertaining only to modes in $S$. 
The characteristic function corresponding to the reduced density operator is 
\begin{equation}
\chi^{(s)}_{\hat{\rho}_S}(\vec{\Lambda}_S)=\mathrm{Tr}[\hat{\rho} D(\vec{\Lambda}_S)]=\chi_{\hat{\rho}}^{(s)}(\vec{\Lambda})\Big|_{\vec{\Lambda}_{\bar{S}}=0},
\end{equation}
which shows that it can be obtained from the total characteristic function by setting the variables pertaining to the complement equal to zero.

In the present case we are interested in $\hat{O}_S=\ketbra{\mathrm{vac}}{\mathrm{vac}}_S$, and it is most convenient to use Eq.~({\ref{ExpectationOComplexS}}) with $s=-1$. For a Gaussian state the $s=-1$ characteristic function is given in Eq.~({\ref{ComplexGaussianChi}}). Using the result above, the characteristic function pertaining only to the modes in $S$ is found by setting $\vec{\Lambda}_{\bar{S}}=0$, meaning we can replace $\sigma$ and $\vec{d}$ with $\sigma_S$ and $\vec{d}_S$, being the covariance matrix and displacement vector retaining rows and columns pertaining to the modes in $S$ only. The $s=+1$ characteristic function for a projection onto the vacuum for the set of modes $S$ is simply 
\begin{equation}
\chi^{(1)}_{\ketbra{\mathrm{vac}}{\mathrm{vac}}_S}(\vec{\Lambda}_S)=1,
\end{equation}
and using this we find
\begin{equation}
P_{\mathrm{off}}(S)=\mathrm{Tr}[\rho \ketbra{\mathrm{vac}}{\mathrm{vac}}_S]=
 \pi^{|S|} W_{\hat{\rho}_S}^{(-1)}(\vec{0})
=\frac{2^{|S|}\exp[-\vec{d}_S^{\dagger}(\sigma_S+\mathds{1}_{2|S|})^{-1}\vec{d}_S]}{\sqrt{\mathrm{det}[\sigma_S+\mathds{1}_{2|S|}]}}.
\end{equation}
Setting $\vec{d}_S=0$ and $\sigma_S=\mathds{1}_{2|S|}$ we find $P_{\mathrm{off}}(S)=1$ as expected.

\end{appendices}

\end{document}